\documentclass[12pt]{article}
\usepackage{a4wide,amssymb,epsfig,authordate1-4,graphicx,color}
\usepackage{tildes,epsfig,graphicx,float,color}
\usepackage[tbtags]{amsmath}
\usepackage{subfigure}

\newcommand{\Pco}{\mbox{$P_o^\mathrm{C}$}}

\DeclareMathOperator{\sech}{sech}

\title{Cellular buckling in stiffened plates\footnote{Accepted for
    publication in \emph{Proc.\ R. Soc.\ A}}}

\author{M. Ahmer Wadee and Maryam Farsi\\~\\
  Department of Civil \& Environmental Engineering,\\
  Imperial College London, London SW7 2AZ, UK}

\date{}

\begin{document}
\maketitle

\begin{abstract}
  An analytical model based on variational principles for a
  thin-walled stiffened plate subjected to axial compression is
  presented. A system of nonlinear differential and integral equations
  is derived and solved using numerical continuation. The results show
  that the system is susceptible to highly unstable local--global mode
  interaction after an initial instability is triggered. Moreover,
  snap-backs in the response showing sequential destabilization and
  restabilization, known as cellular buckling or snaking, arise. The
  analytical model is compared to static finite element models for
  joint conditions between the stiffener and the main plate that have
  significant rotational restraint. However, it is known from previous
  studies that the behaviour, where the same joint is insignificantly
  restrained rotationally, is captured better by an analytical
  approach than by standard finite element methods; the latter being
  unable to capture cellular buckling behaviour even though the
  phenomenon is clearly observed in laboratory experiments.
\end{abstract}

\section{Introduction}

The buckling of thin plates that are stiffened in the longitudinal
direction represents a structural instability problem of enormous
practical significance
\cite{Murray1973,Ronalds89,Sheikh2002,Ghavami2006}. Since
they are primarily used in structures where mass efficiency is a key
design consideration, stiffened plates tend to comprise slender plate
elements that are vulnerable to a variety of different elastic
instability phenomena. In the current work, the classic problem of a
panel comprising a uniform flat plate stiffened by several evenly
spaced blade-type longitudinal stiffeners under axial compression,
made from a linear elastic material, is studied using an analytical
approach. Under this type of loading, wide panels with many stiffeners
can be divided into several struts, each with a single
stiffener. These individual struts are primarily susceptible to a
global (or overall) mode of instability namely Euler buckling, where
flexure about the axis parallel to the main plate occurs once the
theoretical global buckling load is reached. However, when the
individual plate elements of the strut cross-section, namely the main
plate and the stiffener, are relatively thin or slender, elastic local
buckling of these may also occur; if this happens in combination with
the global instability, the resulting behaviour is usually far more
unstable than when the modes are triggered individually
\cite{IUTAM76}.

In the current work, the development of a variational model is
presented that accounts for the interaction between the global Euler
buckling mode and the local buckling mode of the stiffener, such that
the perfect and imperfect elastic post-buckling response can be
evaluated. A system of nonlinear ordinary differential equations
subject to integral constraints is derived and solved using the
numerical continuation package \textsc{Auto-07p} \cite{auto}. It is
indeed found that the system is highly unstable when interactive
buckling is triggered; snap-backs in the response show a sequence of
destabilization and restabilization derived from the mode interaction
and the stretching of the plate surface during local buckling
respectively. This results in a progressive spreading of the initial
localized buckling mode. This latter type of response has become known
in the literature as \emph{cellular buckling} \cite{Hunt2000} or
\emph{snaking} \cite{BurkeKnobloch2007} and it is shown to appear
naturally in the current numerical results. The effect is particularly
strong where the rotational restraint provided at the joint between
the main plate and the stiffener is negligible. As far as the authors
are aware, this is the first time this phenomenon has been found
analytically in stiffened plates undergoing global and local buckling
simultaneously. Similar behaviour has been discovered in various other
mechanical systems such as in the post-buckling of cylindrical shells
\cite{Hunt2003}, the sequential folding of geological layers
\cite{MAW_jmps05}, the buckling of thin-walled I-section beams
\cite{WG2012} and columns \cite{WB2013}.

In previous work, cellular buckling has been captured in physical
tests for some closely related structures that suffer from
local--global mode interaction \cite{Becque2009a,WG2012}; it is
revealed by the buckling elements exhibiting a progressively varying
deformation wavelength. However, by increasing the rotational
stiffness of the aforementioned joint, the snap-backs are moderated
and the equilibrium path shows an initially smoother response with the
snap-backs appearing later in the interactive buckling process. For
these cases, the results from a purely numerical model, formulated
within the commercial finite element (FE) package \textsc{Abaqus}
\shortcite{Abaqus}, are used to validate the analytical model and exhibit
very good correlation, leading to the conclusion that the physics of
the system is captured accurately by the analytical model.

\section{Analytical Model}

\subsection{Modal Description}

\begin{figure}[htbp]
  \centering
  \subfigure[]{\includegraphics[width=150mm]{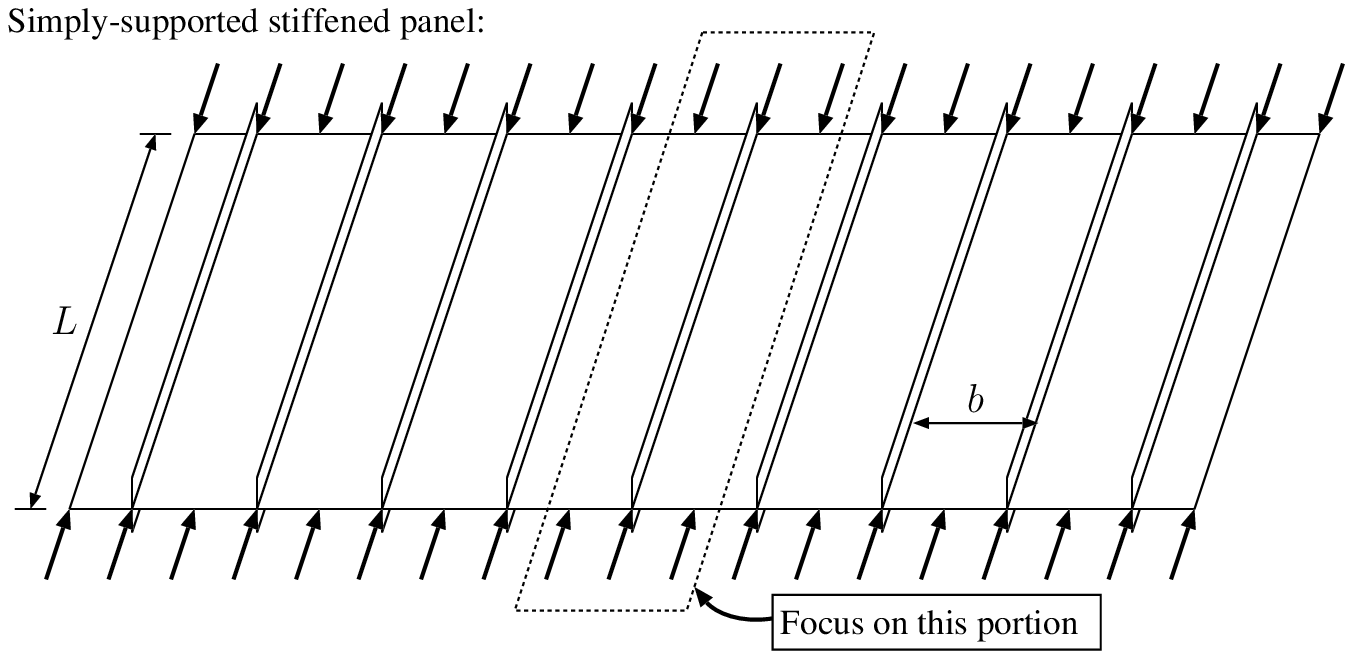}}
  \subfigure[]{\includegraphics[scale=0.8]{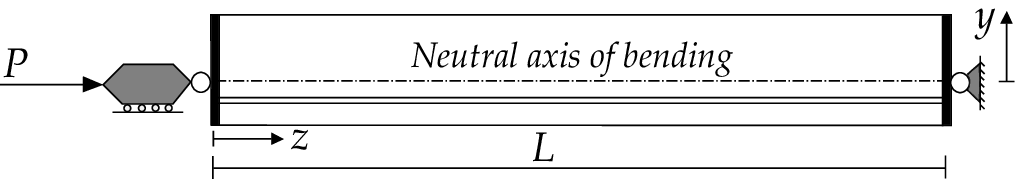}}\qquad
  \subfigure[]{\includegraphics[scale=0.8]{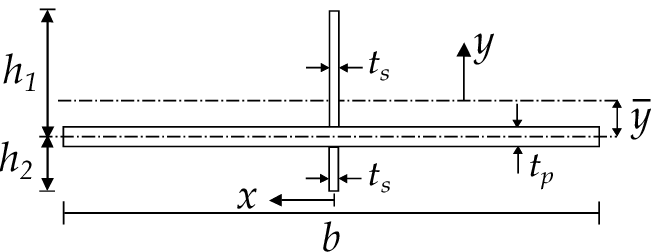}}
  \subfigure[]{\includegraphics[scale=0.8]{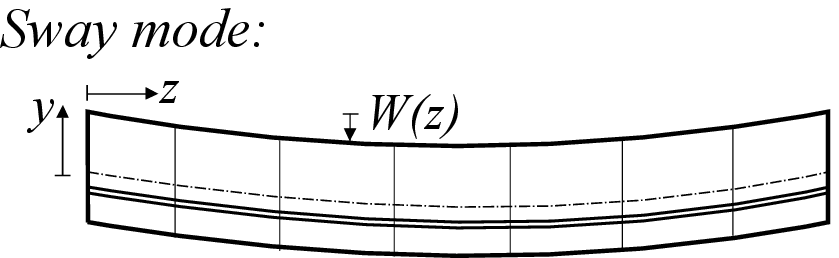}}\qquad
  \subfigure[]{\includegraphics[scale=0.8]{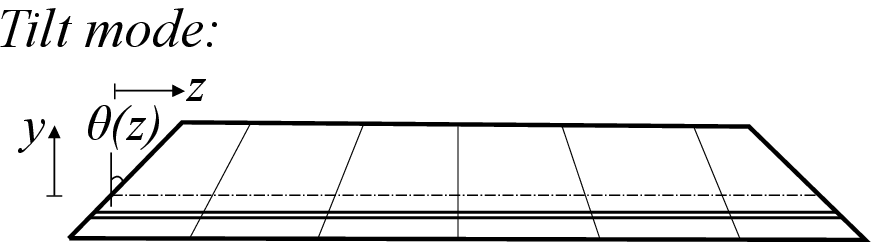}}
  \caption{(a) Axially compressed simply-supported stiffened panel of
    length $L$. (b) Elevation and (c) cross-section of the
    representative panel portion compressed by a force $P$ applied to
    the centroid. (d--e) Sway and tilt components respectively of the
    major axis global buckling mode.}
\label{fig:loadcross}
\end{figure}

Consider a thin-walled simply-supported plated panel that has
uniformly spaced stiffeners above and below the main plate, as shown
in Figure \ref{fig:loadcross}
with the associated coordinate system. The panel length is $L$
and the spacing between the stiffeners is $b$. It is made from a
linear elastic, homogeneous and isotropic material with Young's
modulus $E$, Poisson's ratio $\nu$ and shear modulus
$G=E/[2(1+\nu)]$. A portion of the panel that is representative of its
entirety can be isolated as a strut, shown in Figure
\ref{fig:loadcross}(b--c), since the transverse bending curvature of
the panel during buckling would be relatively small, particularly if
$L \ll n_s b$, where $n_s$ is the number of stiffeners in the
panel. The main plate (width $b$ and thickness $t_p$) has upper and
lower stiffeners (heights $h_1$ and $h_2$ respectively with equal
thickness $t_s$) connected to it. The strut is loaded by an axial
force $P$ that is applied to the centroid of the
cross-section. Presently, however, although the model is formulated
for the general case, the numerical results focus on the practically
significant case where the stiffeners are only connected to one side
of the panel, \emph{i.e.}\ where $h_2=t_p/2$.

Moreover, the geometries are chosen such that global (Euler) buckling
about the $x$-axis is the first instability mode encountered by the
strut, before any local buckling in the main plate or stiffener
occurs. The formulation for global buckling is based on small
deflection assumptions, since it is well known that it has an
approximately flat (or weakly stable) post-buckling response
\cite{TH73}. The study is primarily concerned with global buckling
forcing the stiffener to buckle locally shortly after or even
simultaneously.  It has been shown in several works
\cite{HW1998,WYT2010,WG2012,WB2013} that shear strains need to be
included to model the local--global mode interaction analytically. For
thin-walled components, rather than, for instance, soft core materials
used in sandwich structures \cite{WYT2010}, Timoshenko beam theory
gives sufficiently accurate results \cite{WG2012}. This can be applied
to global flexural buckling within the framework of two degrees of
freedom, known as ``sway'' and ``tilt'' in the literature
\cite{HW1998}. These are associated with functions describing the
lateral displacement $W$ and the angle of inclination $\theta$
respectively defining the appropriate kinematics, as shown in
Figure~\ref{fig:loadcross}(c). From linear theory, it can be shown
that $W(z)$ and $\theta(z)$ can be represented by the following
expressions:
\begin{equation}
  W(z)=-q_s L \sin \frac {\pi z}{L}, \quad
  \theta(z) = q_t \pi \cos\frac{\pi z}{L}.
\end{equation}
The quantities $q_s$ and $q_t$ are the generalized coordinates of the
sway and tilt components of global buckling respectively. The
corresponding shear strain $\gamma_{yz}$ during bending is given by
the following expression:
\begin{equation}
  \gamma_{yz} = \frac{\D W}{\D z} + \theta = -\left(q_s-q_t\right)\pi
  \cos \frac{\pi z}{L}.
\end{equation}
Of course, if standard Euler--Bernoulli bending were used then
$\gamma_{yz}$ would be zero and $q_s$ would be equal to $q_t$. The
initial global buckling displacement is assumed to put the stiffener
into extra compression and it therefore becomes vulnerable to local
buckling. The direction of global buckling is crucially important; if
the stiffener goes into extra compression and buckles locally it can
lead to excessive deflection. This can induce plasticity within the
stiffener, leading to the \emph{tripping} phenomenon that can lead to
catastrophic failure of panels \cite{Ronalds89,Butler2000}; currently,
however, linear elasticity is assumed throughout.

\begin{figure}[htb]\center
\subfigure[]{\includegraphics[scale=0.7]{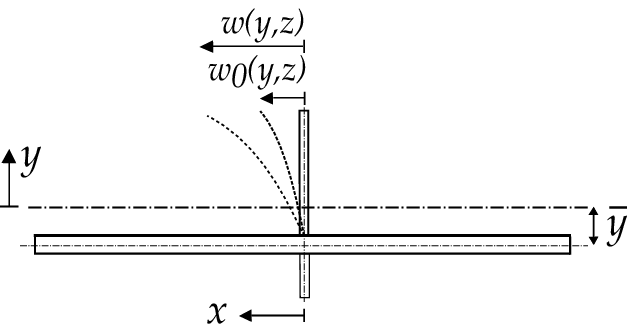}}\quad
\subfigure[]{\includegraphics[scale=0.7]{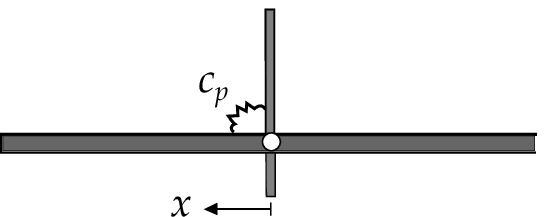}}\quad
\subfigure[]{\includegraphics[scale=0.7]{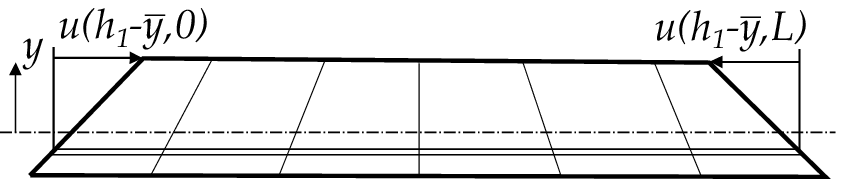}}
\subfigure[]{\includegraphics[scale=0.85]{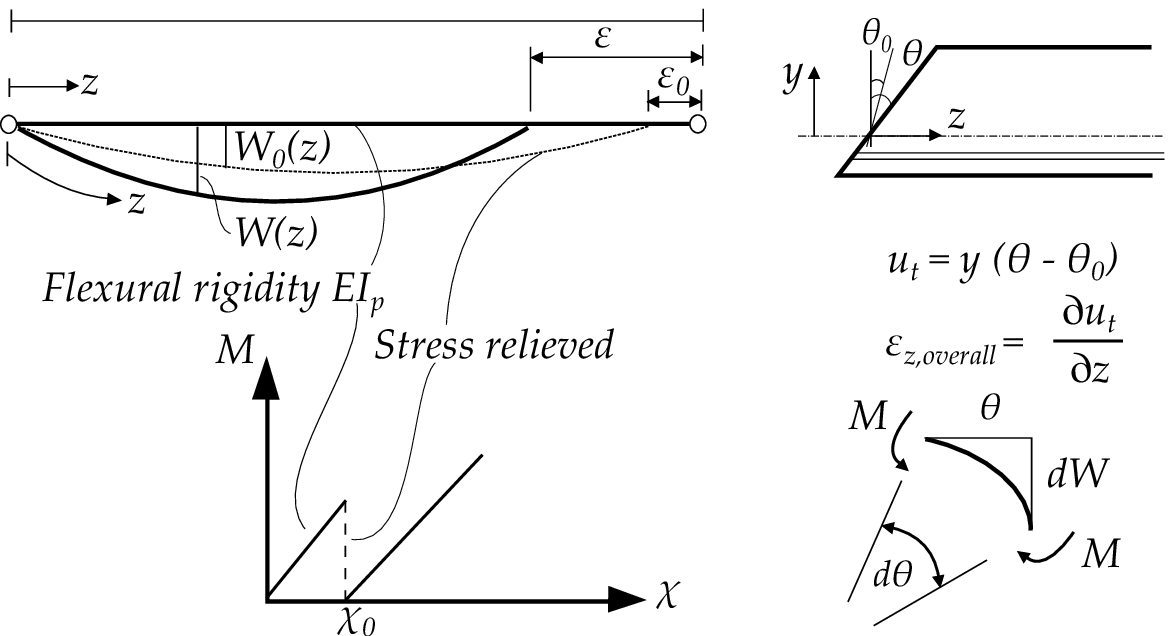}}
\caption{(a) Local out-of-plane displacement $w(y,z)$ and an
  additional displacement $w_0(y,z)$ that represents an initial
  imperfection in the stiffener geometry; (b) a pinned connection with
  an additional rotational spring of stiffness $c_p$ to model joint
  fixity; (c) the local in-plane displacement $u(y,z)$. (d)
  Introduction of the global imperfection functions $W_0$ and
  $\theta_0$; for the consideration of the local imperfection, $w_0$
  would replace $W_0$ and apply only to the stiffener.}
\label{fig:local}
\end{figure}

The stiffener is taken to be a flat plate (blade-type) and the local
kinematics of it require careful consideration. A linear distribution
in $y$ for the local in-plane displacement $u(y,z)$ is assumed owing
to the application of Timoshenko beam theory. The tips of the
stiffeners have free edges but a rotational spring with stiffness
$c_p$ is included to model the resistance to rotation about the
$z$-axis that the joint provides between the stiffeners and the main
plate. Hence, if $c_p=0$, the stiffeners are effectively pinned with
the main plate. In the generic case ($c_p > 0$), however, the assumed
out-of-plane displacement along the width of the stiffener $w(y,z)$
can be approximated by a function that is a summation of trigonometric
and polynomial terms.  Figure~\ref{fig:local}
shows the form of the functions $w$ and $u$ with the algebraic
expressions being given thus:
\begin{equation}
   u(y,z) = Y(y) u(z), \quad
   w(y,z) = f(y) w(z),
\end{equation}
where $Y(y)=(y+\bar{y})/h_1$ and
\begin{equation}
  f(y) = B_0 + B_1 Y + B_2 Y^2 + B_3 Y^3 + B_4 \sin \left( \pi Y \right),
\end{equation}
with $B_0$, $B_1$, $B_2$, $B_3$ and $B_4$ being constants evaluated by
solving for the appropriate boundary conditions for the stiffener. The
choice of shape for $f$ reflects the possibility of the slope at the
junction between the main plate and the stiffener being small in the
less compressed zone of the stiffener. For $y=-\bar{y}$, at the
junction between stiffener and the main plate, the conditions are:
\begin{equation}
  w(\bar{y},z)=0, \quad 
  D\frac{\partial^2}{\partial y^2} w(\bar{y},z)
  = c_p \frac{\partial}{\partial y} w(\bar{y},z),
\end{equation}
and for $y=h_1-\bar{y}$, the free-end, the conditions are:
\begin{equation}
  D\frac{\partial^2}{\partial y^2} w(h_1-\bar{y},z)=0, \quad
  D\frac{\partial^3}{\partial y^3} w(h_1-\bar{y},z)=0,
\end{equation}
where $\bar{y}$ is the location of the neutral-axis of bending
measured from the centre line of the main plate, expressed thus:
\begin{equation}
  \bar{y} = \frac{t_s\left[ h_1^2-h_2^2 \right]}{2
    \left[\left(b-t_s\right)t_p + \left(h_1+h_2\right) t_s \right]}.
\end{equation}
By applying the boundary conditions, the function for the deflected
shape of the stiffener $w(y,z)$ is found to be:
\begin{equation}
  w(y,z) = \left\{ Y - S_4 \frac{\pi^3}{6} \left[ 2 Y - 3 Y^2 + Y^3
      - \frac{6}{\pi^3} \sin \left( \pi Y \right) \right] \right\} w(z),
\end{equation}
where:
\begin{equation}
  S_4 = \left\{\pi \left[(D \pi^2)/(c_p h_1) + \pi^2/3 - 1\right] \right\}^{-1}.
\end{equation}

In the current study, both perfect and imperfect panels are
considered. A displacement $W_0$ corresponding to $W$ and a rotation
$\theta_0$ corresponding to $\theta$ are introduced; the respective
expressions being:
\begin{equation}
  W_0(z) = -q_{s0} L \sin\frac{\pi z}{L}, \quad
  \theta_0 = q_{t0} \pi \cos\frac{\pi z}{L},
\end{equation}
with $q_{s0}$ and $q_{t0}$ being the amplitudes defining the initial
global imperfection.  A local imperfection $w_0(y,z)=f(y)w_0(z)$, see
Figure \ref{fig:local}(a), is also introduced, the form of which is
derived from a first order approximation of a multiple scale
perturbation analysis of a strut on a softening foundation
\cite{WHW97}:
\begin{equation}
  w_0(z) = A_0 \sech \left[\frac{\alpha
      \left(z-\eta\right)}{L}\right] \cos \left[\frac{\beta \pi
      \left(z-\eta \right)}{L}\right],
  \label{eq:localimp}
\end{equation}
where $z = [0,L]$ and $w_0$ is symmetric about $z=\eta$. This form for
$w_0$ has been shown to be representative for local--global mode
interaction problems in the literature \cite{Wadee2000}. It enables
the study of periodic and localized imperfections, the latter being
shown to be relatively more severe for related problems
\cite{delam2,asymmsand}.

\subsection{Total Potential Energy}

The governing equations of equilibrium are derived from variational
principles by minimizing the total potential energy $V$ of the
system. The total potential energy comprises the contributions of
global and local strain energies of bending, $U_{bo}$ and $U_{bl}$
respectively, the strain energy stored in the ``membrane'' of the
stiffener $U_m$ arising from axial and shear stresses, and the work
done by the external load $P\mathcal{E}$. The effects of the initial
imperfections are evaluated by assuming that the imperfect forms,
given by the functions $W_0$ and $w_0$, are stress-relieved. This
implies that the elemental moment $M$ and hence the bending energy
both drop to zero in the unloaded state, see Figure
\ref{fig:local}(d). The global bending energy accounts for the main
plate and is hence given by:
\begin{equation}
  U_{bo} = \frac12 EI_p \int_0^L \left( \ddot{W}-\ddot{W}_0
  \right)^2 \D z=\frac12 EI_p \int_0^L \left(q_s-q_{s0}\right)^2
  \frac{\pi^4}{L^2}\sin^2\frac{\pi z}{L}\D z,
\label{eq:ob}
\end{equation}
where dots represent differentiation with respect to $z$ and
$I_p=(b-t_s)t_p^3/12+(b-t_s)t_p \bar{y}^2$ is the second moment of
area of the main plate about the global $x$-axis. The local bending
energy $U_{bl}$, accounting for the stiffener only, is given thus:
\begin{equation}
\begin{aligned}
  U_{bl} & = \frac{D}{2} \int_0^L
  \int_{-\bar{y}}^{h_1-\bar{y}} \left\{ \left[
  \frac{\partial^2(w-w_0)}{\partial
    z^2}+\frac{\partial^2(w-w_0)}{\partial y^2} \right]^2 \right. \\
  & \left.
    \qquad - 2\left(1-\nu \right) \left[\frac{\partial^2 (w-w_0)}{\partial
      z^2} \frac{\partial^2(w-w_0)}{\partial
      y^2}-\left(\frac{\partial^2(w-w_0)}{\partial z \partial y}
    \right)^2 \right] \right\} \D y \D z, \\
  & = \frac{D}{2} \int_0^L \left[\{ f^2 \}_y \left(\ddot{w}-
      \ddot{w}_0 \right)^2+\bigl\{ f''^2 \bigr\}_y
    (w-w_0)^2 +2 \nu \bigl\{ f f'' \bigr\}_y
    (w-w_0)(\ddot{w}-\ddot{w_0}) \right. \\
  & \left.
    \qquad +2(1-\nu) \bigl\{ f'^2 \bigr\}_y (\dot{w}-\dot{w_0})^2 \right] \D z,
\end{aligned}
\end{equation}
\noindent
where primes denote differentiation with respect to $y$, $D$ is the
plate flexural rigidity given by $Et_s^3/[12(1-\nu^2)]$ and:
\begin{equation}
  \label{eq:braces}
  \{F(y)\}_y = \int_{-\bar{y}}^{h_1-\bar{y}} F(y) \D y,
\end{equation}
where $F$ is an example function. The compressive side of the panel
only contributes to the local bending energy once global buckling
occurs.  The membrane energy $U_m$ is derived from considering the
direct strains ($\varepsilon$) and shear strains ($\gamma$) in the
stiffener. The global buckling distribution for the longitudinal
strain $\varepsilon_z$ can be obtained from the tilt component of
displacement from the global mode, thus:
\begin{equation}
  \varepsilon_{z,\mathrm{global}} = y\frac{\Dx}{\Dx z} \left(\theta -
    \theta_0 \right)
  = -y\left(q_t-q_{t0}\right)\frac{\pi^2}{L}\sin\frac{\pi z}{L}.
\end{equation}
The local mode contribution is based on von K\'arm\'an plate theory. A
purely in-plane compressive strain $\Delta$ is also included. The
direct strain for the main plate is given by: $\varepsilon_{zp}=
-\Delta$. The combined expressions for the direct strains for the top
and bottom stiffeners $\varepsilon_{zt}$ and $\varepsilon_{zb}$
respectively, including local and global buckling are:
\begin{equation}
\begin{aligned}
    \varepsilon_{zt} & = -y \left(q_t-q_{t0}\right)\frac{\pi^2}{L} \sin
  \frac{\pi z}{L}-\Delta + \frac{\partial u}{\partial z} + \frac12
  \left(\frac{\partial w}{\partial z} \right)^2 - \frac12
  \left(\frac{\partial w_0}{\partial z} \right)^2, \\
  & = -y \left(q_t-q_{t0}\right)\frac{\pi^2}{L} \sin \frac{\pi z}{L} -
  \Delta + \left(\frac{y+\bar{y}}{h_1}\right) \dot{u} + \frac12
  \{f^2\}_y \left(\dot{w}^2-\dot{w}_0^2 \right),
\end{aligned}
\end{equation}
\begin{equation}
  \varepsilon_{zb} = -y \left(q_t-q_{t0}\right)\frac{\pi^2}{L} \sin
  \frac{\pi z}{L}-\Delta.
\end{equation}
The complete expression for the membrane strain energy $U_m$ is given
by:
\begin{equation}
\begin{aligned}
  U_m & = U_d + U_s \\
  & = \frac12 \int_0^L \int_{-t_s/2}^{t_s/2} \left[
    \int_{-\bar{y}}^{h_1-\bar{y}}\left( E \varepsilon_{zt}^2 + G
      \gamma_{yzt}^2 \right) \Dx y + \int_{-(h_2+\bar{y})}^{-\bar{y}}
    \left( E \varepsilon_{zb}^2 + G \gamma_{yzb}^2 \right) \Dx y
  \right] \Dx x \D z,
\end{aligned}
\end{equation}
where $U_d$ is the contribution from direct strains, which is given by
the terms multiplied by the Young's modulus $E$; whereas $U_s$, the
contribution arising from the shear strains, is given by the terms
multiplied by the shear modulus $G$. The transverse component of the
strain $\varepsilon_y$ is neglected since it has been shown that it
has no effect on the post-buckling stiffness of a long plate with
three simply-supported edges and one free edge \cite{Koiter1976}. The
total direct strain energy $U_d$ is therefore:
\begin{equation}
\begin{aligned}
  U_d = \frac12 E & t_s \int_0^L \biggl\{ \frac13\left[
    \left(h_1-\bar{y} \right)^3 + \left(h_2+\bar{y} \right)^3 \right]
  \left(q_t-q_{t0}\right)^2 \frac{\pi^4}{L^2} \sin^2 \frac{\pi z}{L}
  \\
  & + \Delta^2 \left(h_1+h_2 \right) + \left[ \left(h_1-\bar{y}
    \right)^2 - \left(h_2+\bar{y} \right)^2 \right] \Delta
  \left(q_t-q_{t0} \right) \frac{\pi^2}{L} \sin \frac{\pi z}{L} \\
  & + h_1 \left[\frac13 {\dot{u}}^2 + \frac{1}{4h_1}\{f^4\}_y
    \left(\dot{w}^2 - \dot{w}_0^2 \right)^2 + \biggl\{\frac{Y
      f^2}{h_1} \biggr\}_y \dot{u}
    \left(\dot{w}^2-\dot{w}_0^2 \right)\right] \\
  & - \left(q_t-q_{t0}\right) \frac{h_1 \pi^2}{L} \sin \frac{\pi z}{L}
  \left[ \left(\frac23 h_1-\bar{y} \right) \dot{u} + \frac{1}{h_1}
    \{yf^2\}_y \left( \dot{w}^2-\dot{w}_0^2 \right)
  \right] \\
  & - h_1 \Delta \left[ \dot{u} + \frac{1}{h_1}\{f^2\}_y
    \left(\dot{w}^2-\dot{w}_0^2 \right) \right] + \left(
    \frac{t_p}{t_s} \right)(b-t_s) \Delta^2 \biggr\} \D z.
\end{aligned}
\end{equation}
The shear strain energy $U_s$ requires the shear strain $\gamma_{yz}$,
which is also modelled separately for the compression and the tension
side of the stiffeners. The general expression for the shear strain
$\gamma_{yzi}$ in the stiffeners is thus:
\begin{equation}
  \gamma_{yzi} = \frac{\Dx}{\Dx z} \left(W - W_0\right) +
  \left(\theta-\theta_0\right) + \frac{\partial u}{\partial y} +
  \frac{\partial w}{\partial z}\frac{\partial w}{\partial y} -
  \frac{\partial w_0}{\partial z}\frac{\partial w_0}{\partial y},
\end{equation}
and the expressions for the top and bottom stiffeners are given
respectively:
\begin{equation}
\begin{aligned}
  \gamma_{yzt} & = -\left[
    \left(q_s-q_{s0}\right)-\left(q_t-q_{t0}\right)\right] \pi \cos
  \frac{\pi z}{L} +\frac{u}{h_1} + \bigl\{f f' \bigr\}_y
  (w\dot{w}-w_0\dot{w_0}),\\
  \gamma_{yzb} & =-\left[ \left(q_s-q_{s0}\right) - \left(q_t - q_{t0}
    \right)\right] \pi \cos \frac{\pi z}{L},
\end{aligned}
\end{equation}
hence, the expression for $U_s$ is:
\begin{equation}
\begin{aligned}
  U_s = \frac12 G & t_s \int_0^L \biggl\{ \left[ \left(q_s-q_{s0}
    \right)-\left(q_t-q_{t0} \right)\right]^2 {\pi}^2 \cos^2 \frac{\pi
    z}{L} \left(h_1+h_2 \right)\\
  & + \frac{1}{h_1} \left[ u^2 + h_1 \bigl\{ \left(f f' \right)^2
    \bigr\}_y \left( w\dot{w}-w_0\dot{w}_0 \right)^2+ 2 \bigl\{ f f'
    \bigr\}_y u \left( w\dot{w}-w_0\dot{w}_0 \right) \right] \\
  & - \left[ \left(q_s-q_{s0} \right)-\left(q_t-q_{t0} \right)\right]
  \biggl[ 2u + 2\bigl\{ f f' \bigr\}_y
  (w\dot{w}-w_0\dot{w}_0 )\biggr] \pi \cos \frac{\pi z}{L} \biggr\}\D
  z.
\end{aligned}
\end{equation}
The strain energy stored in the rotational spring connecting the
stiffeners to the main plate is:
\begin{equation}
  U_{sp} = \frac12 c_p  \int_0^L  \left\{ \frac{\partial}{\partial y}
    \left[w(-\bar{y})-w_0(-\bar{y})\right] \right\}^2 \D z
  = \frac12 c_p  \int_0^L  [f'^2 (-\bar{y})] (w-w_0)^2 \D z,
\end{equation}
where $F(-\bar{y})$ means that the example function $F$ is evaluated
at $y=-\bar{y}$.  The final component of $V$ is the work done by the
axial load $P$, which is given by:
\begin{equation}
  P\mathcal{E} = \frac{P}{2} \int_0^L \left[ 2\Delta + q_s^2 \pi^2
    \cos^2 \frac{\pi z}{L} - 2\left(
      \frac{h_2+\bar{y}}{h_1+h_2} \right) \dot{u} \right] \Dx z,
\end{equation}
\noindent
where the total end-displacement $\mathcal{E}$ comprises components
from pure squash and sway from global buckling combined with the local
buckling of the stiffener respectively. The total potential energy $V$
is given by the summation of all the strain energy terms minus the
work done, thus:
\begin{equation}
   V = U_{bo} + U_{bl} + U_m +U_{sp}- P\mathcal{E}.
\end{equation}

\subsection{Variational Formulation}

The governing equilibrium equations are obtained by performing the
calculus of variations on the total potential energy $V$ following a
well established procedure that has been detailed in Hunt and Wadee
\shortcite{HW1998}. The integrand of the total potential energy $V$
can be expressed as the Lagrangian ($\mathcal{L}$) of the form:
\begin{equation}
V=\int_0^L \mathcal{L}\left(\ddot w, \dot{w}, w, \dot {u}, u, z \right) \D z. 
\end{equation}
The first variation of $V$ is given by:
\begin{equation}
  \delta V = \int_0^L \left[ \frac{\partial \mathcal{L}}{\partial \ddot
      w} \delta \ddot w + \frac{\partial \mathcal{L}}{\partial \dot w}
    \delta \dot w + \frac{\partial \mathcal{L}}{\partial w} \delta 
    w + \frac{\partial \mathcal{L}}{\partial \dot u} \delta \dot u +
    \frac{\partial \mathcal{L}}{\partial u} \delta u \right] \D z,
\end{equation}
to determine the equilibrium states, $V$ must be stationary, hence the
first variation $\delta V$ must vanish for any small change in $w$ and
$u$. Since $\delta\ddot w=\Dx(\delta\dot w)/\D z$, $\delta\dot w=\Dx
(\delta w)/\Dx z$ and similarly $\delta \dot u = \Dx (\delta u)/\Dx
z$, integration by parts allows the development of the Euler--Lagrange
equations for $w$ and $u$; these comprise a fourth-order and a
second-order nonlinear differential equation for $w$ and $u$
respectively. To facilitate the solution of the equations within the
package \textsc{Auto-07p}, the variables are rescaled with respect to
the non-dimensional spatial coordinate $\tilde z$, defined as $\tilde
z = 2z/L$. Similarly, the non-dimensional out-of-plane and in-plane
displacements $\tilde{w}$ and $\tilde{u}$ are defined with the
scalings $2w/L$ and $2u/L$ respectively. Note that the scalings
exploit symmetry about the midspan and the equations are hence solved
for half the strut length; this assumption has been shown to be
acceptable for cases where global buckling is critical
\cite{Wadee2000}. The non-dimensional differential equations are thus:
\begin{equation}
  \begin{aligned}
    \tilde{\ddddot{w}} & - \tilde{\ddddot{w_0}} + \frac{L^2}{2
      \{f^2\}_y} \left[\nu \left\{f f'' \right\}_y - (1-\nu) \left\{
        f'^2 \right\}_y \right] \left( \tilde{\ddot{w}} -
      \tilde{\ddot{w}}_{0}\right) + \tilde{k}
    \left(\tilde{w}-\tilde{w_{0}}\right) \\
    & - \tilde{D} \left[ \frac{\left\{ f^4 \right\}_y}{\left\{
          f^2\right\}_y} \left( 3 \tilde{\dot{w}}^2 \tilde{\ddot{w}} -
        \tilde{\ddot{w}} {\dot{w}_{0}}^2 - 2
        \tilde{\ddot{w}}_{0}\tilde{\dot{w}}_{0} \tilde{\dot{w}}
      \right) + \frac {\left\{2 Y f^2 \right\}_y}{\left\{
          f^2\right\}_y} \left( \tilde{\ddot{u}} \tilde{\dot{w}} +
        \tilde{\ddot{w}}
        \tilde{\dot{u}} \right) \right. \\
    & \left.  - 2 \Delta \tilde{\ddot{w}} - 2 \left( q_t-q_{t0}
      \right) \frac{\pi^2}{L} \frac {\left\{ yf^2 \right\}_y}{\left\{
          f^2\right\}_y} \left( \sin \frac{\pi \tilde{z}}{2}
        \tilde{\ddot{w}} + \frac{\pi}{2} \cos \frac{\pi \tilde{z}}{2}
        \tilde{\dot{w}}
      \right) \right] \\
    & - \frac{\tilde{G} L^2 \tilde{w}}{2\left\{ f^2\right\}_y} \biggl[
    \left\{ \left( f f' \right)^2 \right\}_y \left( \tilde{\dot{w}}^2
      + \tilde{w} \tilde{\ddot{w}} - \tilde{\dot{w}}_{0}^2 -
      \tilde{w_{0}} \tilde{\ddot{w}}_{0}
    \right) + \frac1h_1\left\{f f' \right\}_y \tilde{\dot{u}}  \\
    & + \left[ \left( q_s-q_{s0} \right)- \left( q_t -q_{t0} \right)
    \right] \frac{\pi^2}{L} \left\{f f' \right\}_y \sin \frac{\pi
      \tilde{z}}{2} \biggr]= 0,
\end{aligned}
\label{eq:wdddd}
\end{equation}
\begin{equation}
\begin{aligned}
  \tilde{\ddot{u}} & - \frac34 \frac{\tilde{G}}{\tilde{D}} \psi
  \biggl\{ \psi \left[ \tilde{u}+ \left\{ ff' \right\}_y \left(
      \tilde{w} \tilde{\dot{w}} - \tilde{w}_{0} \tilde{\dot{w}}_{0}
    \right)\right] - 2 \pi \left[
    \left(q_s-q_{s0}\right)-\left(q_t-q_{t0}\right) \right] \cos
  \frac{\pi \tilde{z}}{2} \biggr\} \\
  & - \left\{ \frac{3 Y}{h_1} f^2 \right\}_y \left(
    \tilde{\dot{w}}\tilde{\ddot{w}} + \tilde{\dot{w}}_{0}
    \tilde{\ddot{w}}_{0} \right) + \frac12 \left(q_t-q_{t0}\right)
  \pi^3 \left( \psi - \frac{3\bar{y}}{2L} \right) \cos \frac{\pi
    \tilde{z}}{2} = 0,
\end{aligned}
\label{eq:udd}
\end{equation}
\noindent
where the rescaled quantities are:
\begin{equation}
  \tilde{D}=Et_sL^2/{8D}, \quad \tilde{G}=Gt_sL^2/{8D}, \quad
  \tilde{k}= \frac{L^4}{16 \left\{f^2 \right\}_y} \biggl[
  \left\{f''^2\right\}_y + c_p f'^2(-\bar{y})/D \biggr],
\end{equation}
with $\tilde{w}_0 = 2 w_0/L$, $\psi=L/h_1$ and $f'^2(-\bar{y})$ is as
described previously.  Equilibrium also requires the minimization of
the total potential energy with respect to the generalized coordinates
$\Delta$, $q_s$ and $q_t$ leading to the three integral equations in
nondimensional form:
\begin{equation}  
  \begin{split}
    \frac{\partial V}{\partial \Delta} & = \Delta \left[1+
      \frac{h_2}{h_1} + \frac{t_p(b-t_s)}{t_s} \right] -\frac{P}{Et_s
      h_1} + \left(q_t-q_{t0} \right) \frac{\pi}{h_1 L} \left[
      \left(h_1-\bar{y}
        \right)^2 - \left(h_2+\bar{y} \right)^2 \right] \\
    & -\frac14 \int_0^2 \left[ \tilde{\dot{u}} +\frac{1}{h_1} \left\{
        f^2 \right\}_y \left(\tilde{\dot{w}}^2-\dot{w}_{0}^2 \right)
    \right] \Dx \tilde{z} =0,
\end{split}
\label{eq:equil_int_Del}  
\end{equation}
\begin{equation}
  \begin{split}    
  \frac{\partial V}{\partial q_s} & = \pi^2 \left(q_s - q_{s0}\right) +
  \tilde{s} \left[ \left(q_s - q_{s0}\right)-\left(q_t-q_{t0}\right)
  \right] -\frac{PL^2}{EI_p} q_s \\
  & \quad -\frac{\tilde{s}\tilde{\phi}}{2\pi}\int_0^2 \cos
  \frac{\pi\tilde{z}}{2} \left[ \tilde{u}+ \left\{f f' \right\}_y
    \left( \tilde{w} \tilde{\dot{w}} - \tilde{w}_{0}
      \tilde{\dot{w}}_{0} \right) \right] \Dx \tilde{z} =0,
  \end{split}
\label{eq:equil_int_qs}
\end{equation}
\begin{equation}
  \begin{split}
  \frac{\partial V}{\partial q_t} & = \pi^2 \left(q_t-q_{t0}\right) +
  \tilde{\Gamma_3} \Delta -\tilde{t} \left[
    \left(q_s-q_{s0}\right)-\left(q_t-q_{t0}\right) \right] - \frac12
  \int_0^2 \biggl\{ \sin \frac{\pi \tilde{z}}{2} \left[
    \tilde{\Gamma_1}\tilde{\dot{u}} \right.\\
  & \quad \left.+ \tilde{\Gamma_2} \left( \tilde{\dot{w}}^2
      - \tilde{\dot{w}}_0^2 \right) \right]
  - \frac{\tilde{t} \tilde{\phi}}{\pi} \cos \frac{\pi \tilde{z}}{L} \left[
    \tilde{u}+ \left\{ f f' \right\}_y
    \left(\tilde{w}\tilde{\dot{w}} - \tilde{w_{0}}
    \tilde{\dot{w}}_{0} \right) \right]\biggr\}
  \Dx \tilde{z} = 0,
  \end{split}  
  \label{eq:equil_int_qt}
\end{equation}
where the rescaled quantities are:
\begin{equation}
\begin{aligned}
  \tilde{\Gamma_1} &=
  \frac{Lh_1\left(2h_1-3\bar{y}\right)}{\left(h_1-\bar{y}\right)^3 +
    \left(h_2+\bar{y}\right)^3},
  \quad \tilde{\Gamma_2}=\frac{3L\left\{ yf^2\right\}_y}
  {\left(h_1-\bar{y}\right)^3+\left(h_2+\bar{y}\right)^3}, \\
  \tilde{\Gamma_3} &=\frac{6L \left[ \left(h_1-\bar{y}\right)^2
      -\left(h_2+\bar{y} \right)^2 \right]}{\pi \left[
      \left(h_1-\bar{y}\right)^3 +\left(h_2+\bar{y} \right)^3
    \right]}, \quad \tilde{\phi} = \frac{L}{h_1+h_2}, \\
  \tilde{s} &= \frac{G t_s (h_1+h_2) L^2}{EI_p},\quad
  \tilde{t} =\frac{3GL^2(h_1+h_2)}{E\left[\left(h_1-\bar{y}\right)^3
      +\left(h_2+\bar{y} \right)^3 \right]}.
\end{aligned}
\end{equation}

Since the strut is an integral member, Equations
(\ref{eq:equil_int_qs})--(\ref{eq:equil_int_qt}) provide a
relationship linking $q_s$ and $q_t$ before any interactive buckling
occurs, \emph{i.e.}\ when $w=u=0$. This link is assumed to hold also
between $q_{s0}$ and $q_{t0}$, which acts as a simplification by
reducing the number of global imperfection amplitude parameters to
one; this relationship is given by:
\begin{equation}
q_{s0}=\left( 1+ \pi^2/\tilde{t} \right)q_{t0}.
\end{equation}
The boundary conditions for $\tilde{w}$ and $\tilde{u}$ and their
derivatives are for simply-supported conditions at $\tilde{z}=0$ and
for symmetry conditions at $\tilde{z}=1$:
\begin{equation}
\label{eq:bc_w}
\tilde{w}(0) = \tilde{\ddot{w}}(0) = \tilde{\dot{w}}(1) =
\tilde{\dddot{w}}(1) = \tilde{u}(1) = 0,
\end{equation}
with a further condition from matching the in-plane strain:
\begin{equation}
  \label{eq:bc_ud}
  \frac13 \tilde{\dot{u}}(0) + \frac12 \left\{ \frac{Y}{h_1}
    f^2 \right\}_y \left[
    \tilde{\dot{w}}^2(0)-\dot{w}_{0}^2(0) \right] - \frac12 \Delta +
  \frac{P}{Et_sh_1} \left( \frac{h_2+\bar{y}}{h_1+h_2}\right) = 0.
\end{equation}
It is worth noting that the terms in braces in Equations
(\ref{eq:wdddd})--(\ref{eq:bc_ud}) are integrated with respect to the
original definition of $y$.

Linear eigenvalue analysis for the perfect strut
($q_{s0}=q_{t0}=A_0=0$) is conducted to determine the critical load
for global buckling $\Pco$. By considering that the Hessian matrix
$\mathbf{V}_{ij}$, thus:
\begin{equation}
\mathbf{V}_{ij}=\left[
\begin{array}{cc}
  \frac{\partial^2 V}{\partial q_s^2} & \frac{\partial^2
    V}{\partial q_s \partial q_t} \\
  \frac{\partial^2 V}{\partial q_t \partial q_s} & \frac{\partial^2
    V}{\partial q_t^2}
\end{array}
\right]
\end{equation}
at the critical load is singular, in conjunction with the pre-buckling
condition $q_s=q_t=w=u=0$, the critical load for global buckling is obtained:
\begin{equation}
  \Pco = \frac{\pi^2 EI_p}{L^2} \left[1+ \frac{\tilde{s}}{
      \pi^2+ \tilde{t}} \right].
  \label{eq:pc}
\end{equation}
\noindent
If the limit $G \rightarrow \infty$ is taken, a primary assumption in
Euler--Bernoulli bending theory, $\Pco$ reduces to the expected
classical Euler column buckling load.

\section{Numerical study}
\label{sec:numerics}

The full system of equilibrium equations are difficult to solve
analytically. The continuation and bifurcation software
\textsc{Auto-07p} is thus used; it has been shown in
previous work \cite{WG2012} to be an ideal tool to solve the
equations for this kind of mechanical system.  The solver is adept at
locating bifurcation points and tracing branching paths as model
parameters are varied. To demonstrate this, an example set of
geometric and material properties are chosen thus: $L=5000~\mm$,
$b=120~\mm$, $t_p = 2.4~\mm$, $t_s=1.2~\mm$, $h_1=38~\mm$,
$h_2=t_p/2$, $E = 210~\mathrm{kN/mm^2}$, $\nu = 0.3$.  The global
critical load $\Pco$ can be calculated using Equation (\ref{eq:pc}),
whereas the local buckling critical stress $\sigma_l^\mathrm{C}$ can
be evaluated using the well-known formula $\sigma_l^\mathrm{C}=k_p
D\pi^2/(b^2t)$, where the coefficient $k_p$ depends on plate boundary
conditions; limiting values being $k_p=0.426$ and $k_p=1.247$ for a
long stiffener connected to the main plate with one edge free and the
edge defining the junction between the stiffener and the main plate
being taken to be pinned or fixed respectively.  For the panel
selected, Table \ref{tab:plpc}
\begin{table}[tbh]
\centering
\begin{tabular}{cccc}
  $\sigma_o^\mathrm{C}~(\mathrm{N/mm^2})$ &
  $\sigma_{l,s}^\mathrm{C}~(\mathrm{N/mm^2})$ &
  $\sigma_{l,p}^\mathrm{C}~(\mathrm{N/mm^2})$ &
  Critical mode\\
  \hline
  $4.89 $ & $80.22$ & $373.28$ & Global\\
  \hline
\end{tabular}
\caption{Theoretical values of the global and local critical buckling
  stresses ($\sigma_o^\mathrm{C}$ and $\sigma_l^\mathrm{C}$)
  respectively; subscripts ``$p$'' and ``$s$'' refer to the main plate
  and the stiffener respectively. The expression for
  $\sigma_o^\mathrm{C}=\Pco/A$, where $A$ is the cross-sectional area
  of the strut.}
\label{tab:plpc}
\end{table}
shows that global buckling is critical and that the stiffener could be
the next in line to buckle locally since its critical stress is much
less than that of the main plate.

\subsection{Analytical model results}

\begin{figure}[htb]
\centering
\includegraphics[scale=1.0]{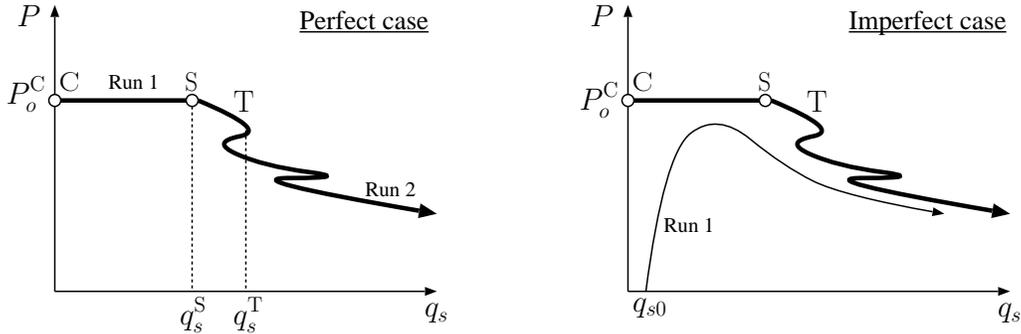}
\caption{Diagrammatic representation of the sequence for computing the
  equilibrium paths for the (left) perfect and (right) imperfect
  cases.}
\label{fig:path}
\end{figure}

To solve the governing equations, the principal parameters used in the
numerical continuation procedure in \textsc{Auto-07p} were
interchangeable although generally $q_s$ was varied for computing the
equilibrium paths for the distinct buckling modes. However, the load
$P$ was used as the principal continuation parameter for computing the
interactive buckling paths. The perfect post-buckling path was
computed first from $\Pco$ and then a sequence of bifurcation points
were detected on the weakly stable post-buckling path; the value of
$q_s$ closest to the critical bifurcation point ($\mathrm{C}$) was
identified as the secondary bifurcation point ($\mathrm{S}$), see
Figure \ref{fig:path}, and denoted as $q_s^\mathrm{S}$.
This triggered interactive buckling and a new equilibrium path was
computed. Typically this reduced the load $P$ with $w$ and $u$
becoming non-trivial as the interaction advanced. On the interactive
buckling path, a progressive sequence of destabilization and
restabilization, exhibited as a series of snap-backs in the
equilibrium paths was observed -- the signature of cellular buckling
\cite{Hunt2000,MAW_jmps05,WG2012,WB2013}. In Figure \ref{fig:path},
the first such snap-back is labelled as $\mathrm{T}$ with $q_s =
q_s^\mathrm{T}$ and the existence of it marks the beginning of the
rapid spreading of the buckling profile from being localized to being
eventually periodic.

\begin{figure}[htbp]
\centering
\subfigure[]{\includegraphics[scale=0.85]{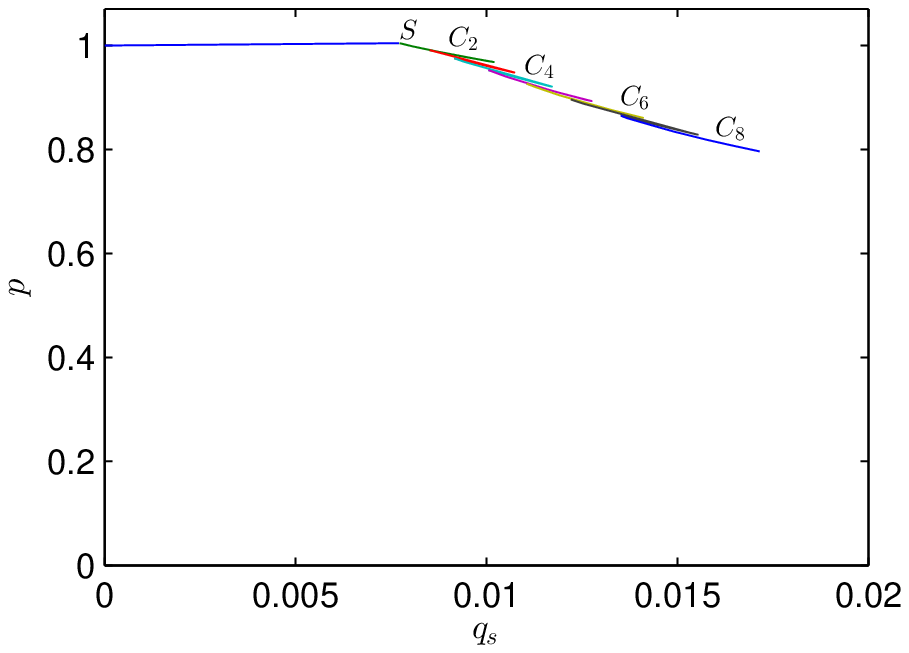}}
\subfigure[]{\includegraphics[scale=0.85]{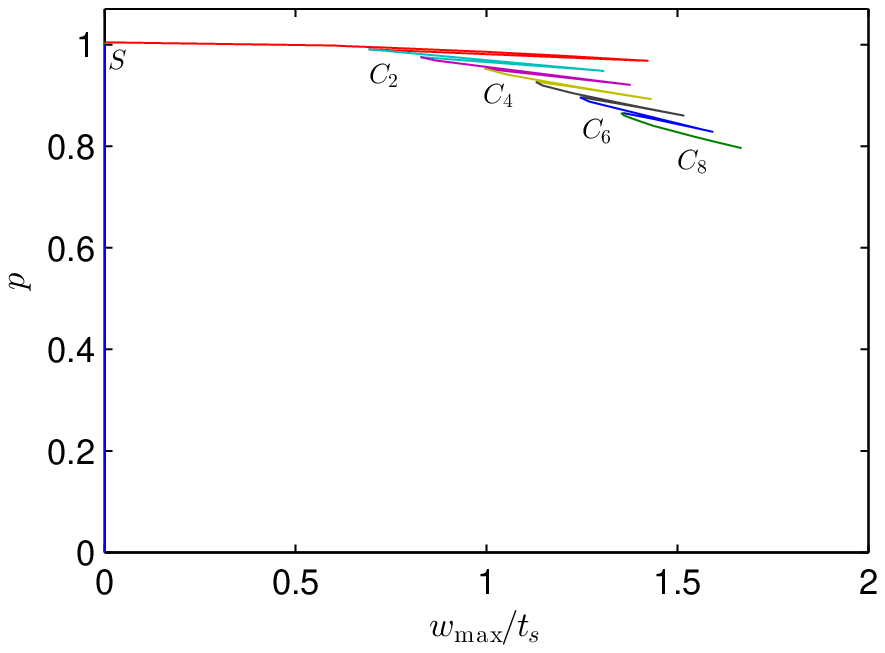}}
\subfigure[]{\includegraphics[scale=0.80]{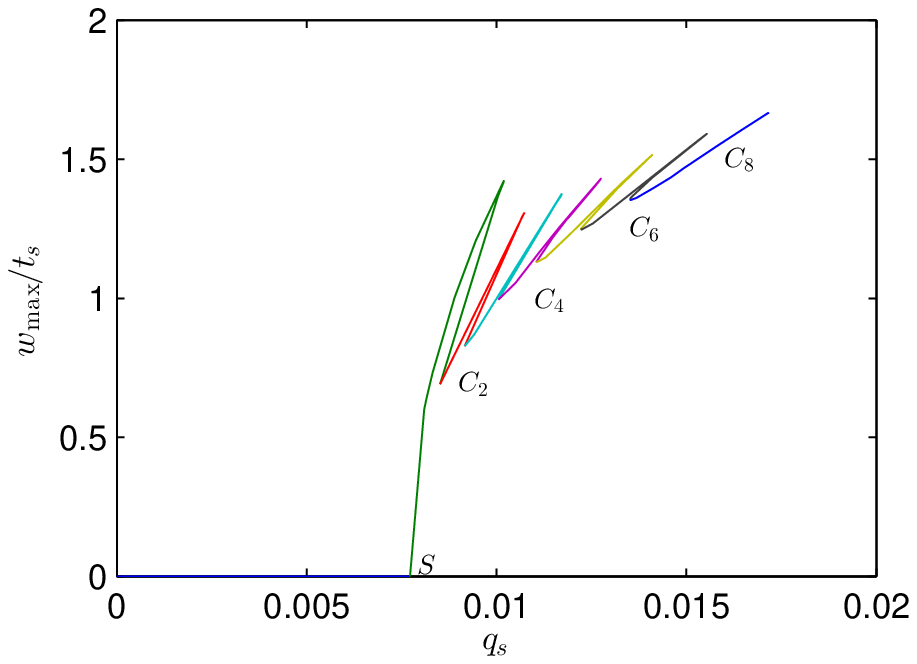}}
\subfigure[]{\includegraphics[scale=0.80]{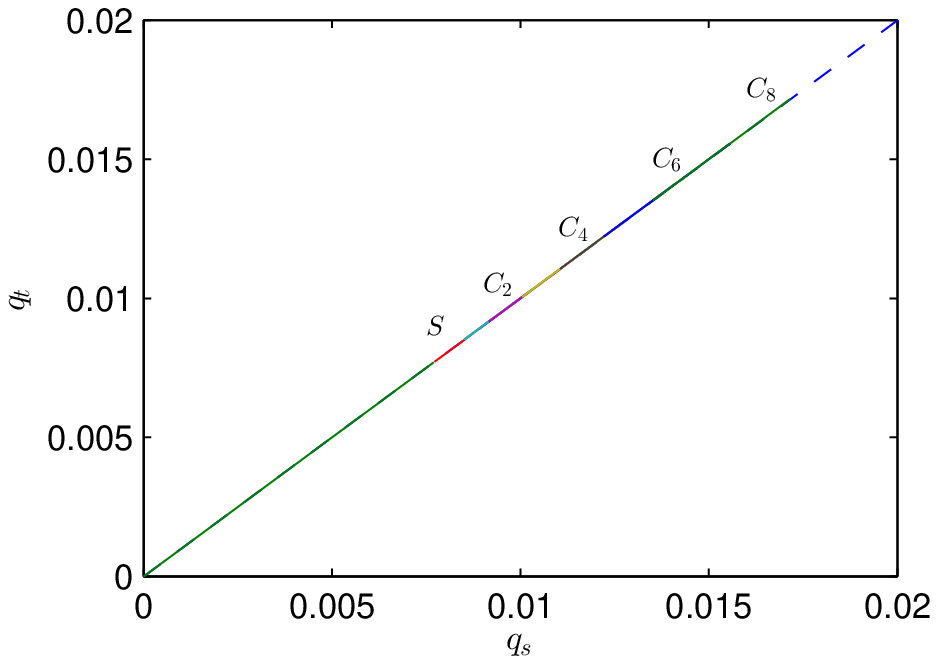}}
\caption{Numerical solutions from the analytical model: equilibrium
  paths. Normalized force $p$ versus the normalized amplitudes of (a)
  global buckling $q_s$ and (b) local buckling of the stiffener
  $w_\mathrm{max}/t_s$. Modal amplitude comparisons: (c) local versus
  global; (d) sway versus tilt.}
\label{fig:pg}
\end{figure}

\begin{figure}[htb]
\centering
\subfigure[]{\includegraphics[scale=0.7]{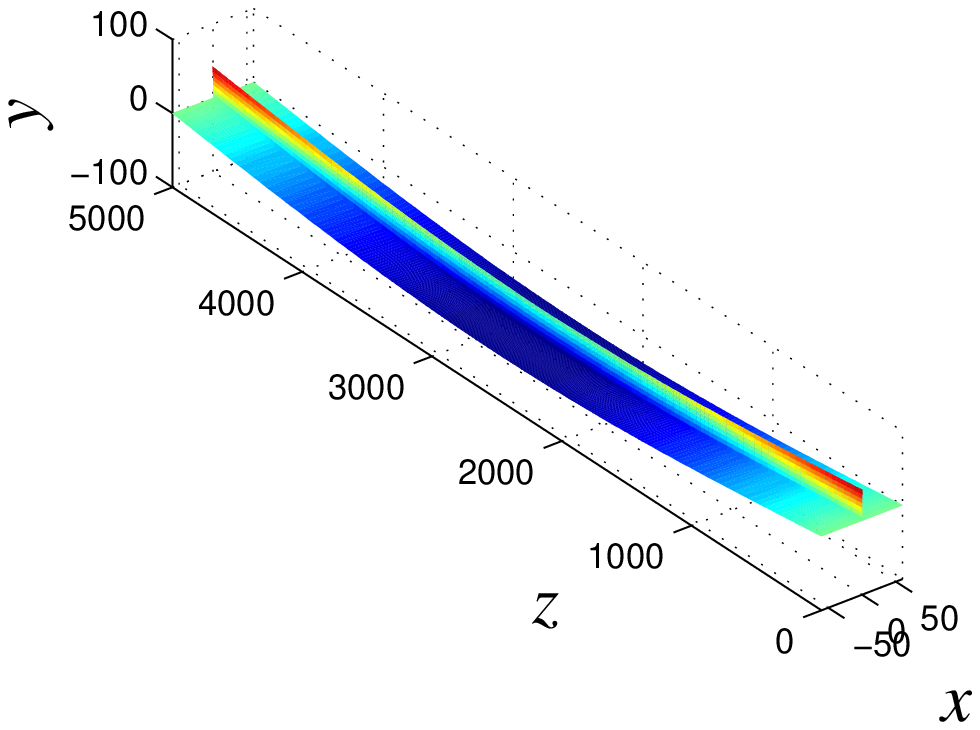}}
\subfigure[]{\includegraphics[scale=0.7]{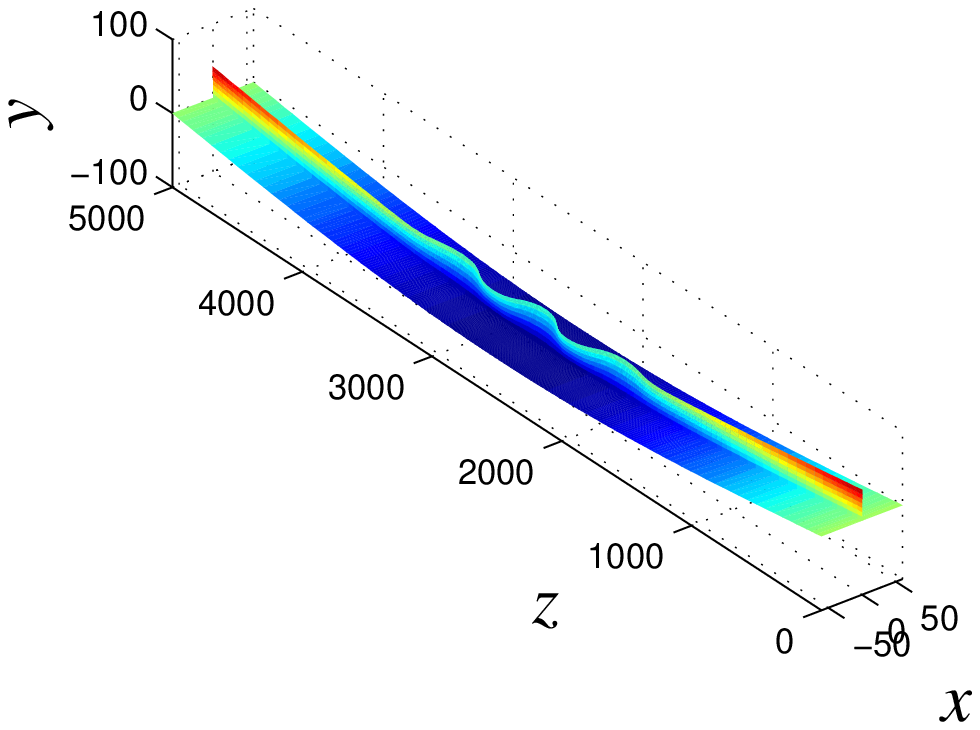}}
\subfigure[]{\includegraphics[scale=0.7]{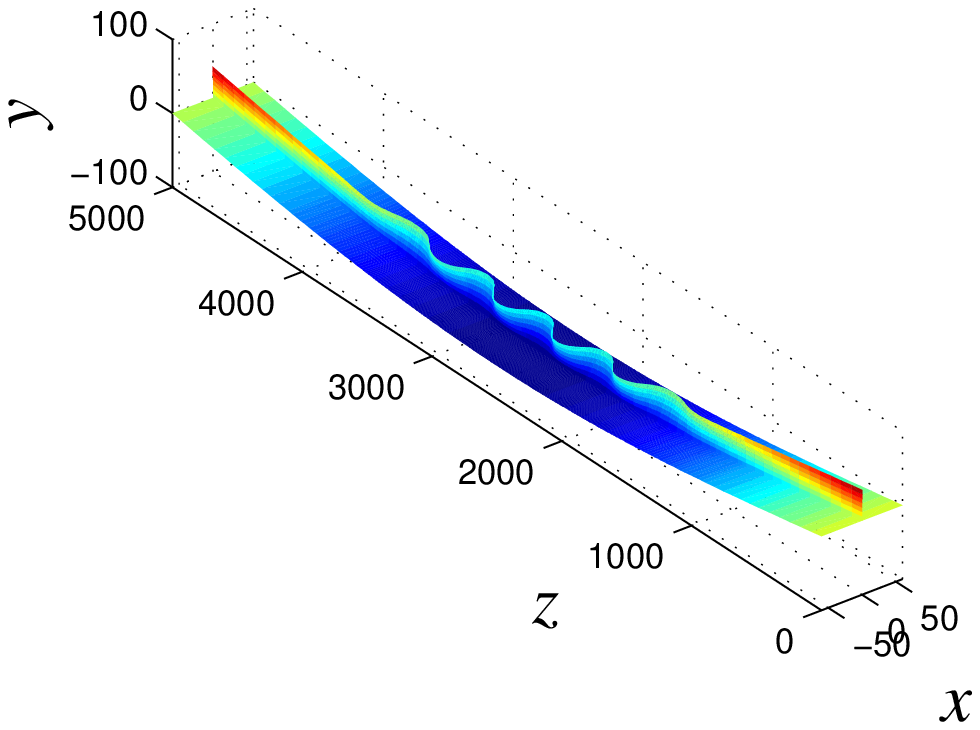}}
\subfigure[]{\includegraphics[scale=0.7]{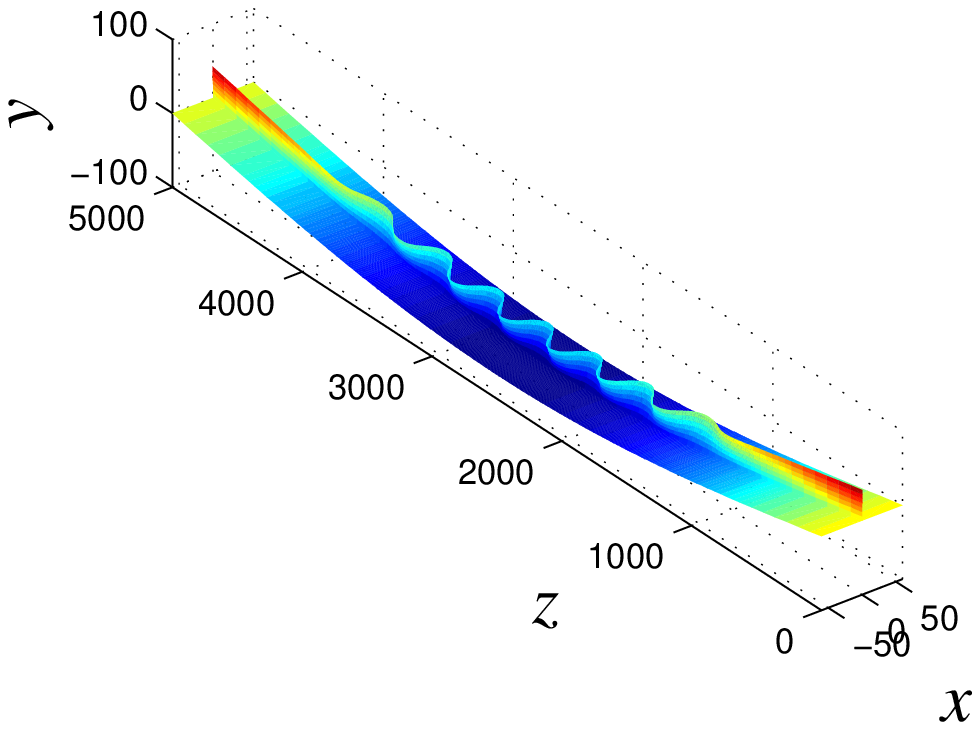}}
\caption{Numerical solutions from the analytical model visualized on
  3-dimensional representations of the strut. (a) Secondary
  bifurcation point $\mathrm{S}~(p=1.00)$, (b) cell $C_4~(p=0.976)$,
  (c) cell $C_6~(p=0.926)$ and (d) cell $C_8~(p=0.865)$. All
  dimensions are in millimetres, but the local buckling displacements
  in the stiffener are scaled by a factor of 5 to aid visualization.}
\label{fig:3dperfect}
\end{figure}

\begin{figure}[htb]
\centering
\includegraphics[scale=0.9]{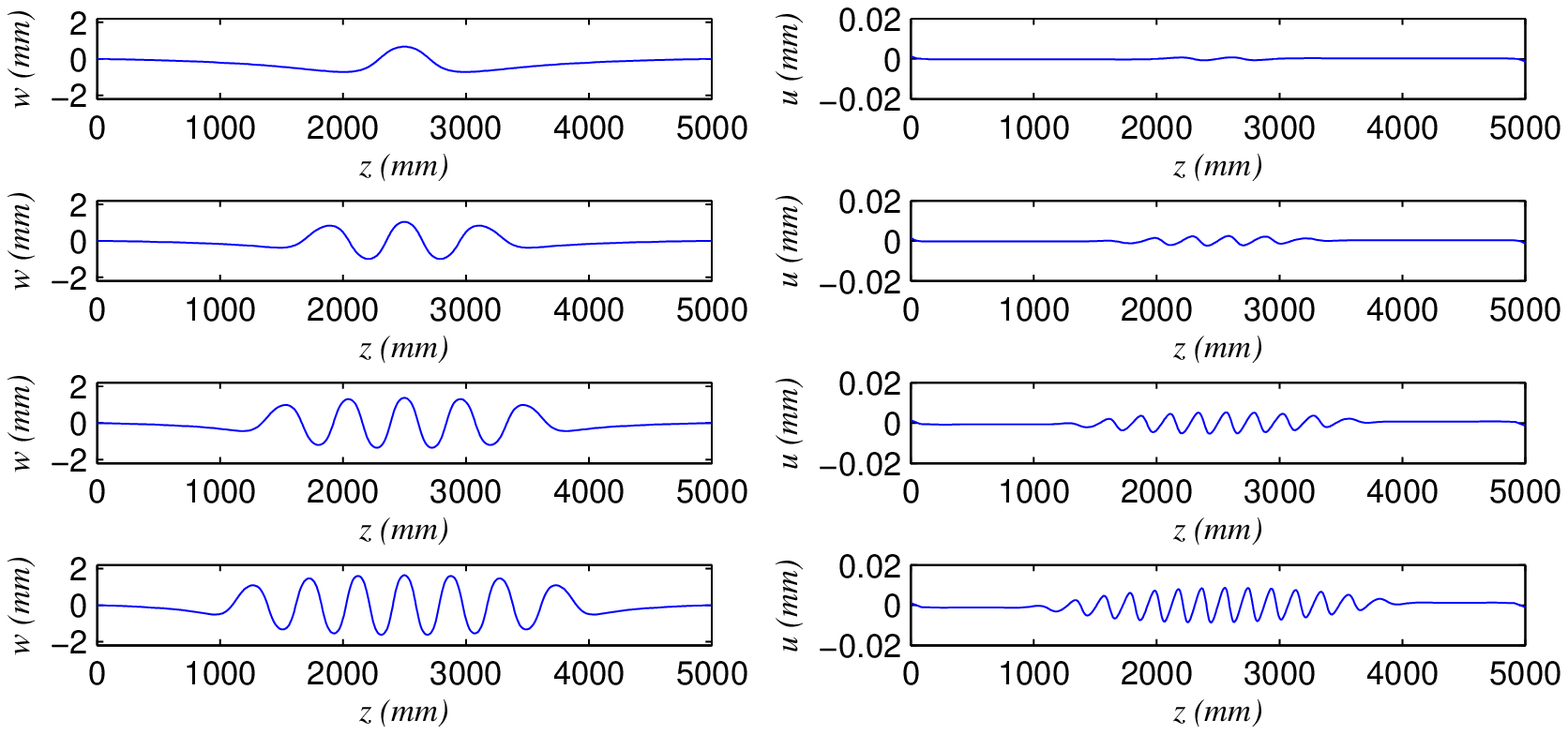}
\caption{Numerical solutions for the local buckling mode out-of-plane
  deflection $w$ (left) and in-plane deflection $u$ (right) showing
  the progressive spreading of the stiffener deflection for cells
  $C_2$, $C_4$, $C_6$ and $C_8$ (top to bottom respectively) of the
  perfect case from the analytical model.}
\label{fig:perfect-wu}
\end{figure}

For the example panel being considered, Figure \ref{fig:pg}
shows equilibrium plots of the normalized axial load $p=P/\Pco$ versus
the normalized buckling amplitudes of (a) the global mode of the strut
and (b) the local mode of the stiffener. The graph in (c) shows the
relative amplitudes of the global and local buckling modes. Finally,
(d) shows the relationship between the sway and tilt amplitudes of
global buckling, which are almost equal; indicating that the shear
strain from the global mode is small but, importantly, not
zero. Figure \ref{fig:3dperfect}
shows a selection of 3-dimensional representations of the deflected
strut that include the components of global buckling ($q_s$ and
$q_t$) and local buckling ($w$ and $u$) for $\mathrm{S}$ and cells
$C_4$, $C_6$, and $C_8$ respectively. Figure \ref{fig:perfect-wu}
illustrates the corresponding progression of the numerical solutions
for the local buckling functions $w$ and $u$. The results for the
example clearly shows cellular buckling with the spiky features in the
graphs in Figure \ref{fig:pg}(a--c). Moreover, the buckling patterns,
as seen in Figure \ref{fig:3dperfect}, clearly show an initially
localized buckle gradually spreading outwards from the panel midspan
with more peaks and troughs.

In the literature \cite{Hunt2000,MAW_jmps05} the Maxwell load
($P^\mathrm{M}$) is calculated for snaking problems, which represents
a realistic lower bound strength for the system. Presently, however,
it is more complex to determine such a quantity because the system
axial load $P$ does not oscillate about a fixed load $P^\mathrm{M}$ as
the deformation increases. This is primarily owing to the fact that,
unlike systems that exhibit cellular buckling, such as cylindrical
shells and confined layered structures \cite{Hunt2000}, there are two
effective loading sources as the mode interaction takes hold: the
axial load $P$, which generally decreases and the sinusoidally varying
(in $z$) tilt generalized coordinate $q_t$, which represents the axial
component of the global buckling mode that generally increases. The
notion of determining the ``body force'', discussed in Hunt and Wadee
\shortcite{HW1998}, could be used as way to calculate the Maxwell
load, but this has been left for future work.

\begin{figure}[htb]
\centering
\mbox{\subfigure[$p=0.99$]{\includegraphics[scale=0.85]{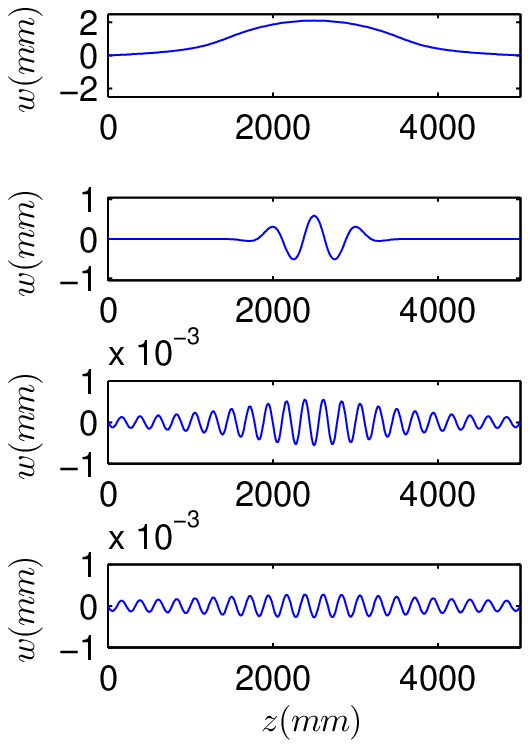}}}
\mbox{\subfigure[$p=0.96$]{\includegraphics[scale=0.85]{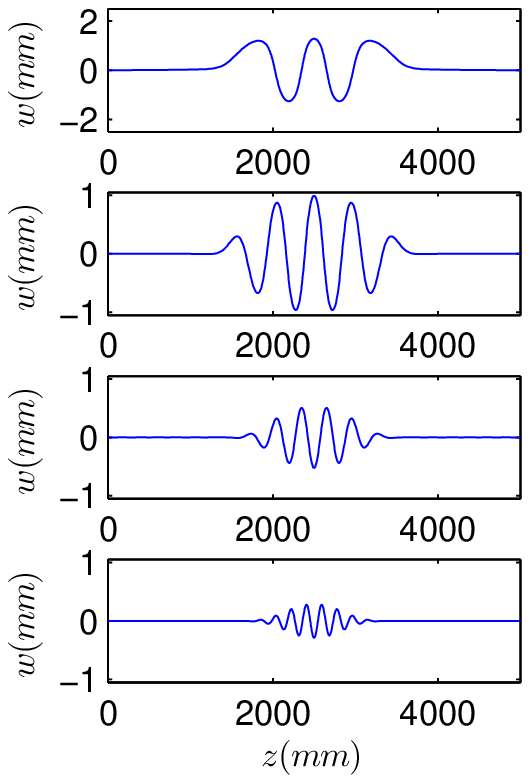}}}
\mbox{\subfigure[]{\includegraphics[scale=0.70]{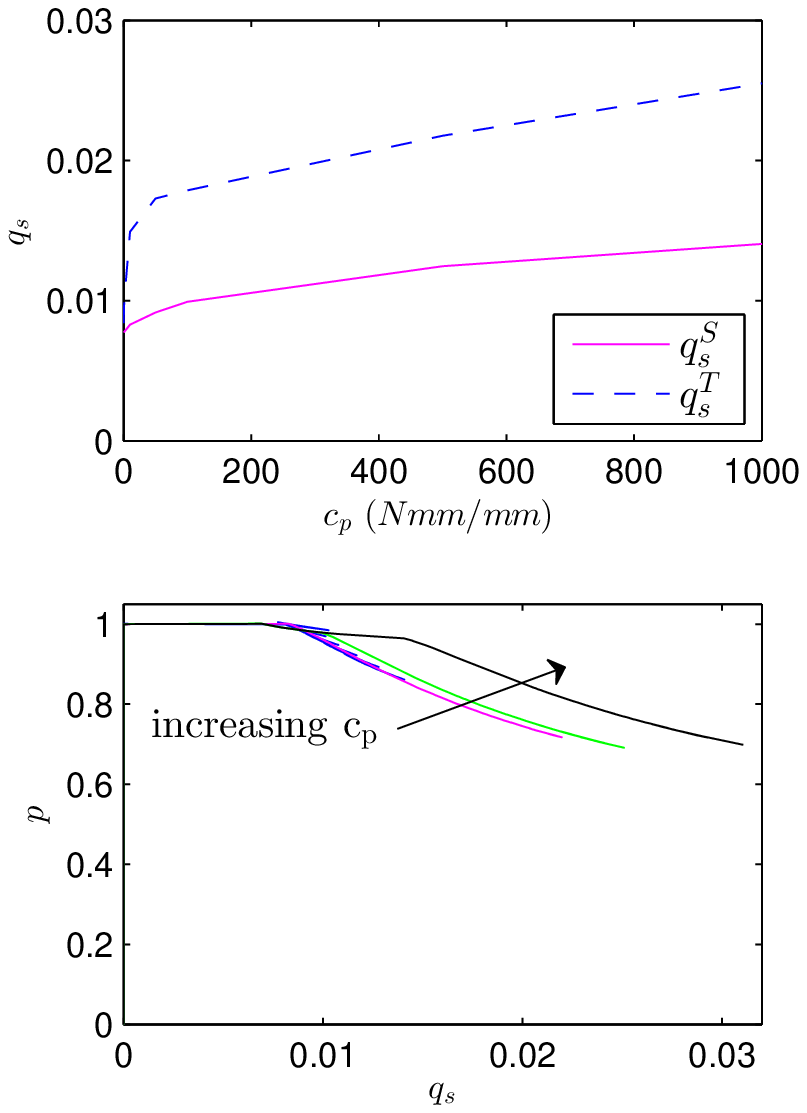}}}
\caption{(a--b) Numerical solutions from the analytical model for the
  local out-of-plane deflection $w$ in the initial interactive
  buckling stage, where cases for $c_p=0,10,100,1000~\mathrm{Nmm/mm}$
  are shown from top to bottom respectively. (c) Curves showing the
  distribution of $q_s^\mathrm{S}$ and $q_s^\mathrm{T}$, and the load
  $p$ versus $q_s$ for the increasing values of $c_p$ given in
  (a--b).}
\label{fig:cp_w}
\end{figure}

Hitherto, the results have been presented for the case where the joint
between the main plate and the stiffener is pinned ($c_p=0$). Figure
\ref{fig:cp_w}
shows how the introduction of the joint stiffness affects the
response. First of all, the value of $\sigma_{l,s}^\mathrm{C}$ is
increased since rotation at the joint is more restrained. This, in
turn, increases the value of $q_s^\mathrm{S}$ and $q_s^\mathrm{T}$ at
different rates. It can be seen that as $c_p$ is increased the initial
interactive mode that emerges from the secondary bifurcation
$\mathrm{S}$ is more localized with a smaller wavelength (Figure
\ref{fig:cp_w}) and resembles the more advanced cells shown in Figure
\ref{fig:perfect-wu} directly. Moreover, it is demonstrated that the
snap-backs occur later in the interactive buckling process.

\section{Validation}

The commercial FE software \textsc{Abaqus} was selected to validate
the results from \textsc{Auto-07p} where appropriate. Four-noded shell
elements with reduced integration were used to model the
structure. Rotational springs were also used along the length to
simulate the boundary conditions at the joint of the stiffener with
the main plate. An eigenvalue analysis was used to calculate the
critical buckling loads and eigenmodes. The nonlinear post-buckling
analysis was performed with the static Riks \shortcite{Riks79} method
with the aforementioned eigenmodes being used to introduce the
necessary geometric imperfection to facilitate this.  For the
numerical example, the strut described in \S\ref{sec:numerics} has a
rotational spring stiffness $c_p=1000~\mathrm{Nmm/mm}$. As discussed
at the end of the previous section, this basically increases the gap
between triggering the interactive buckling mode and the onset of
cellular buckling behaviour since the initial local buckling mode is
naturally more spread through the length; this smoothing and hence
simplification of the response potentially allows for a better
comparison with the analytical model.

\begin{figure}[htbp]
\centering
\subfigure[]{\includegraphics[scale=0.65]{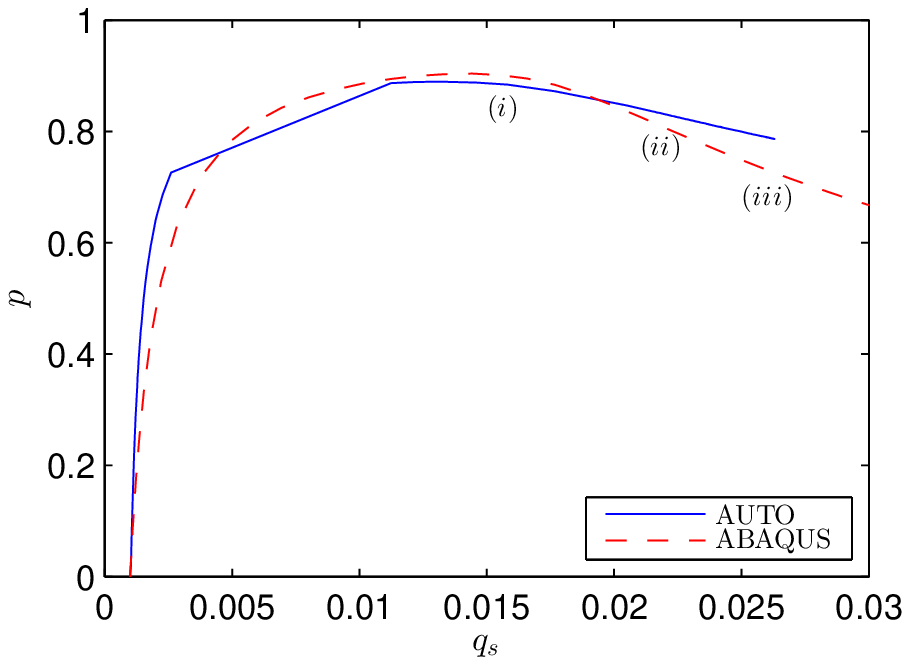}}
\qquad \subfigure[]{\includegraphics[scale=0.65]{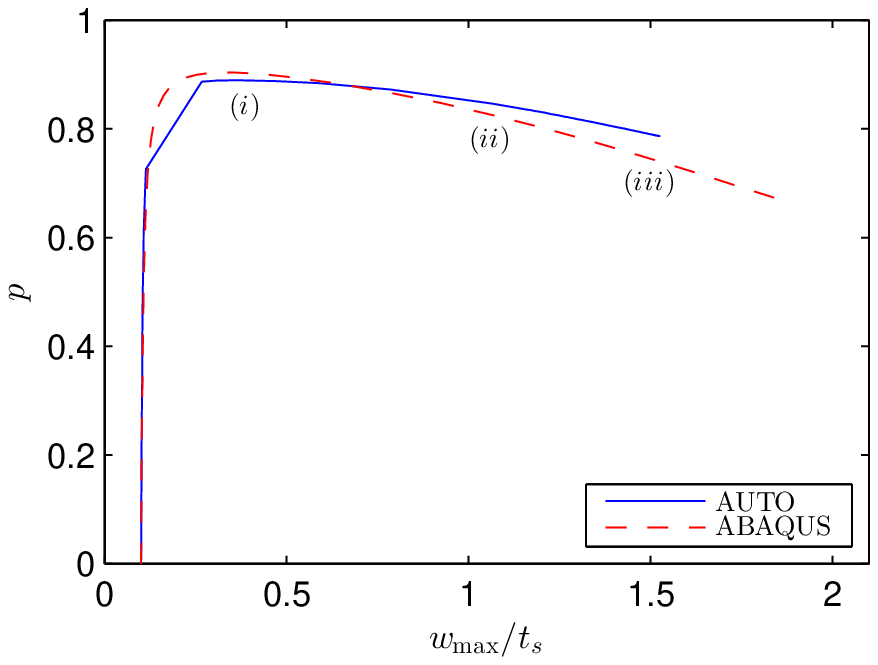}}
\subfigure[]{\includegraphics[scale=0.65]{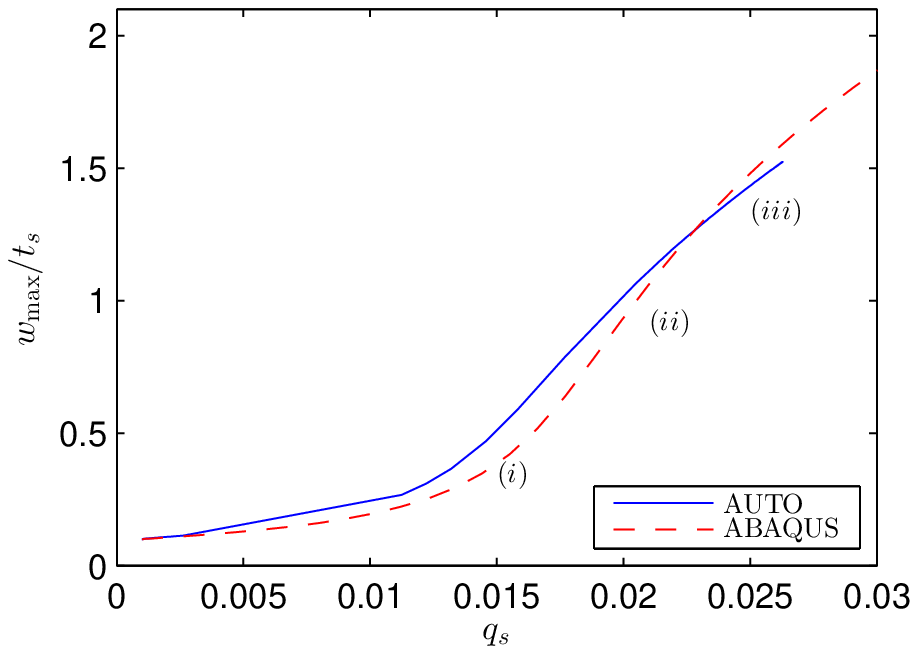}}
\subfigure[]{\includegraphics[scale=0.85]{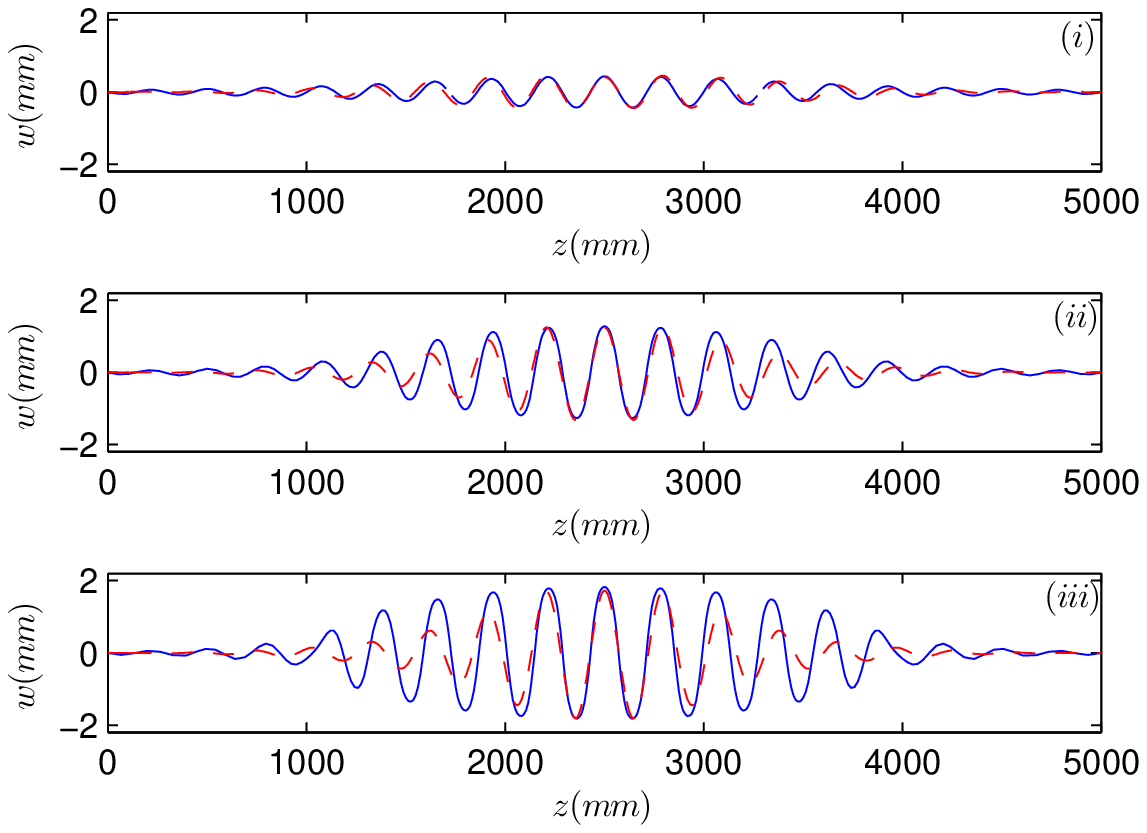}}
\caption{Comparison of the results from \textsc{Auto} (solid lines)
  with \textsc{Abaqus} (dashed lines) solutions. Normalized force $p$
  versus amplitudes of (a) the global mode $q_s$ (b) the local mode
  $w_\mathrm{max}/t_s$. (c) local versus global mode amplitudes; (d)
  local buckling out-of-plane deflections $w$ of the stiffener for the
  points shown in (a--c) defined as (i)--(iii).}
\label{fig:abaqus}
\end{figure}

Linear buckling analysis shows that global buckling is the first
eigenmode and the corresponding critical load and critical stress are
$\Pco=1.64~\kN$ and $\sigma_o^\mathrm{C}=4.95~\mathrm{N/mm^2}$
respectively, which is a difference of approximately $1\%$ from the
value calculated from the analytical model (Table
\ref{tab:plpc}). Figure \ref{fig:abaqus}
shows the comparisons between the analytical results from
\textsc{Auto-07p} and the numerical results from \textsc{Abaqus}. This
is shown for the case with an initial global imperfection ($W_0$)
where $q_{s0}=0.001$ is combined with a local imperfection ($w_0$)
where $A_0=0.12~\mathrm{mm}$, $\alpha=8.0$, $\beta=35$ and $\eta=L/2$,
which matches the initial imperfection from \textsc{Abaqus} for the
analytical model such that a meaningful comparison can be made. The
graphs in (a--b) show the normalized axial load $p$ versus the
buckling amplitudes of the global mode and the local deflection of the
stiffener respectively. The graph in (c) shows the local versus the
global buckling modal amplitudes; all three graphs
show excellent correlation in all aspects of the mechanical response.

\begin{figure}[htbp]
\centering
\subfigure[]{\includegraphics[scale=0.7]{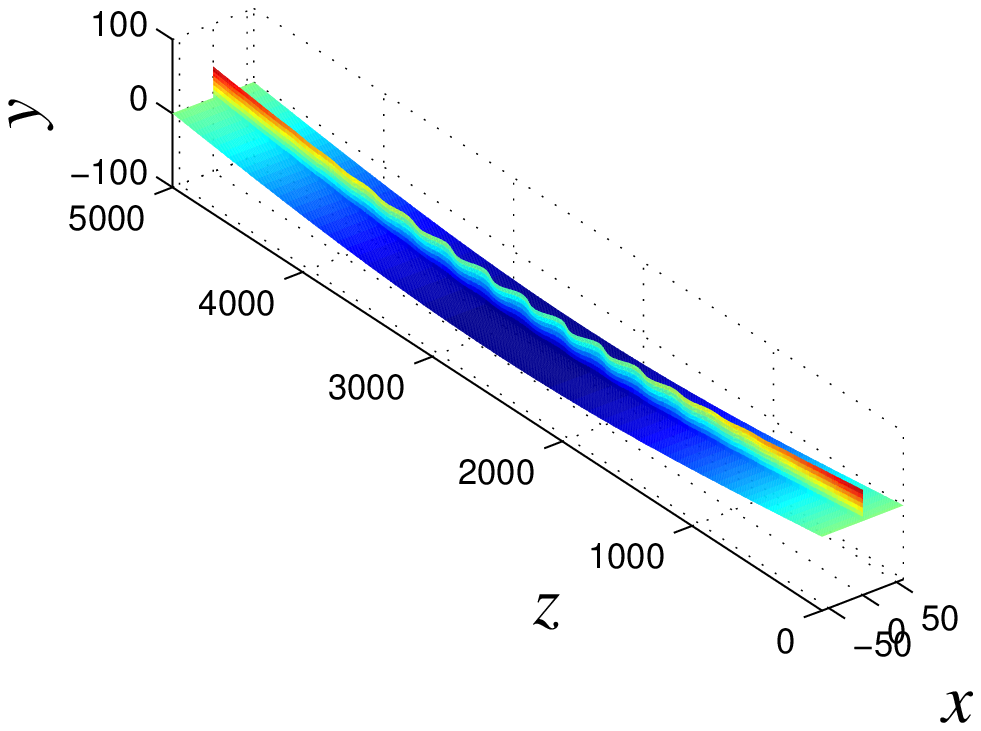}
\quad \includegraphics[scale=0.7]{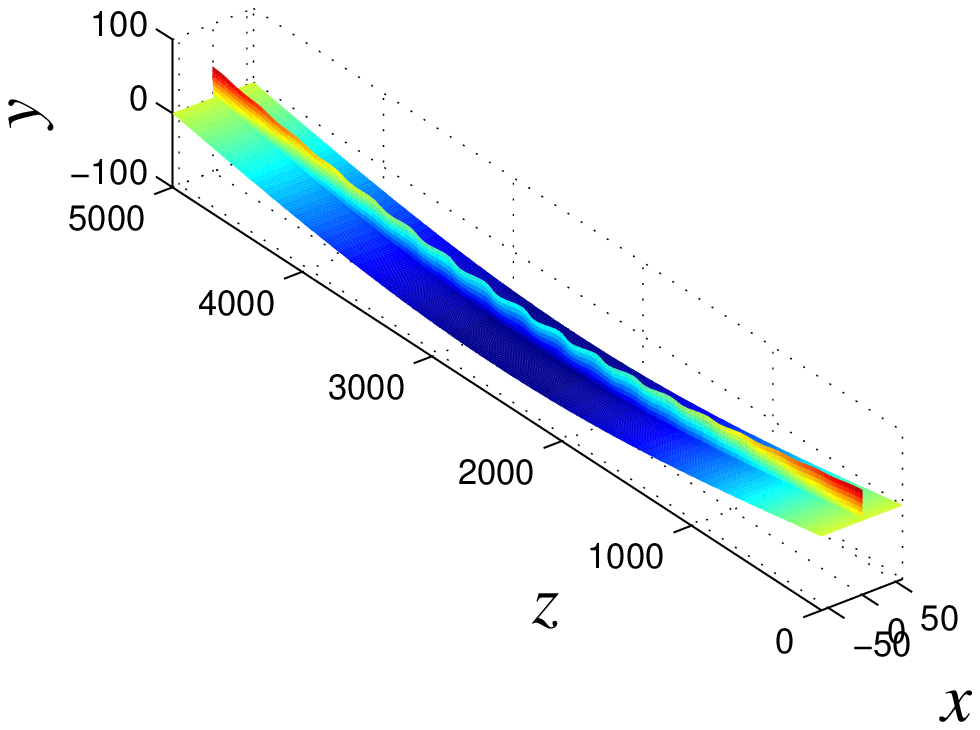}}
\subfigure[]{\includegraphics[scale=0.7]{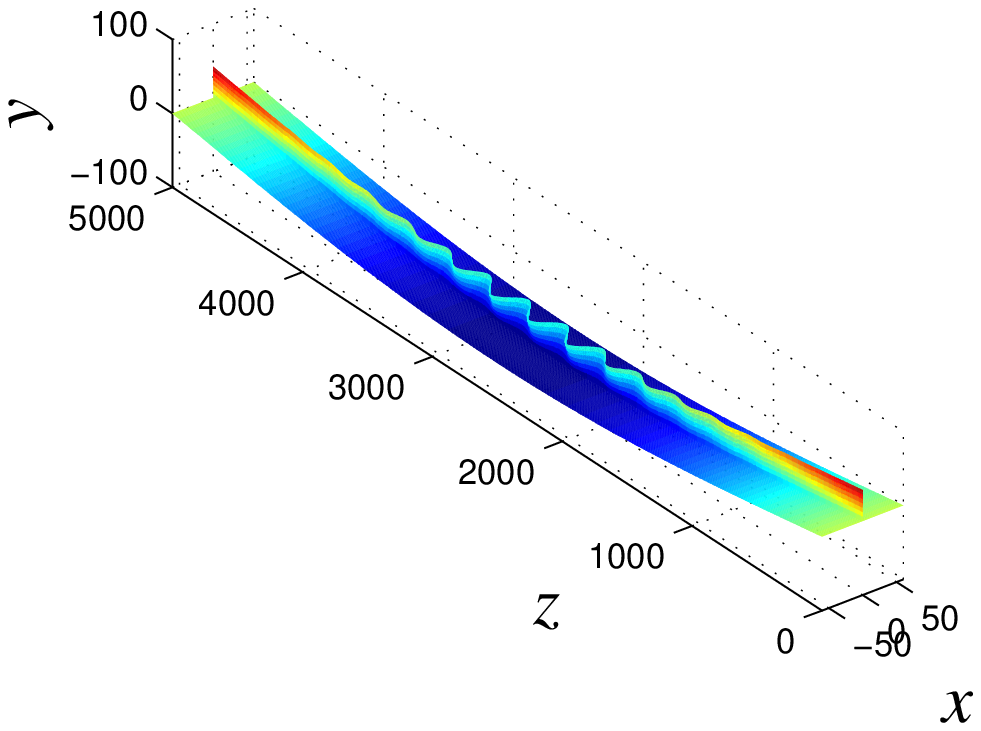}
\quad \includegraphics[scale=0.7]{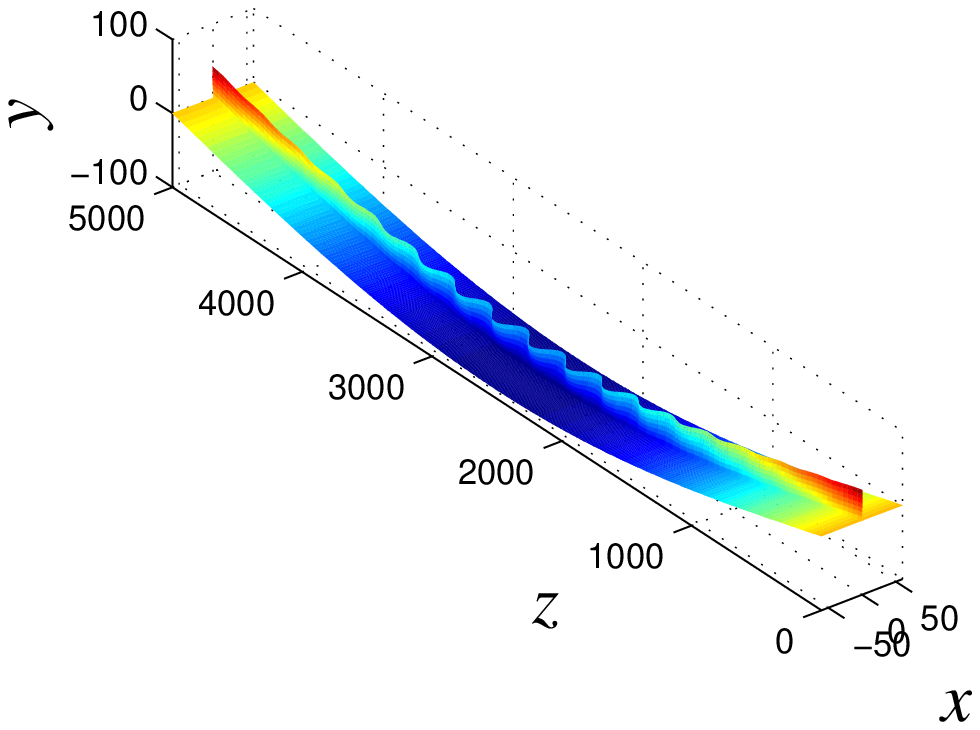}}
\subfigure[]{\includegraphics[scale=0.7]{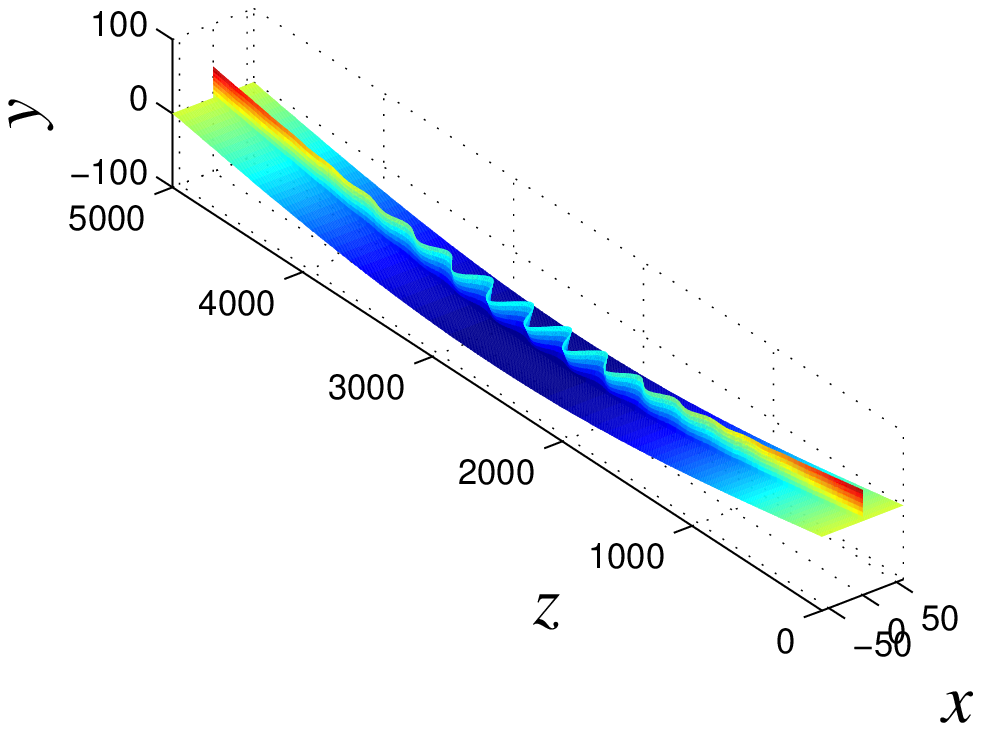}
\quad \includegraphics[scale=0.7]{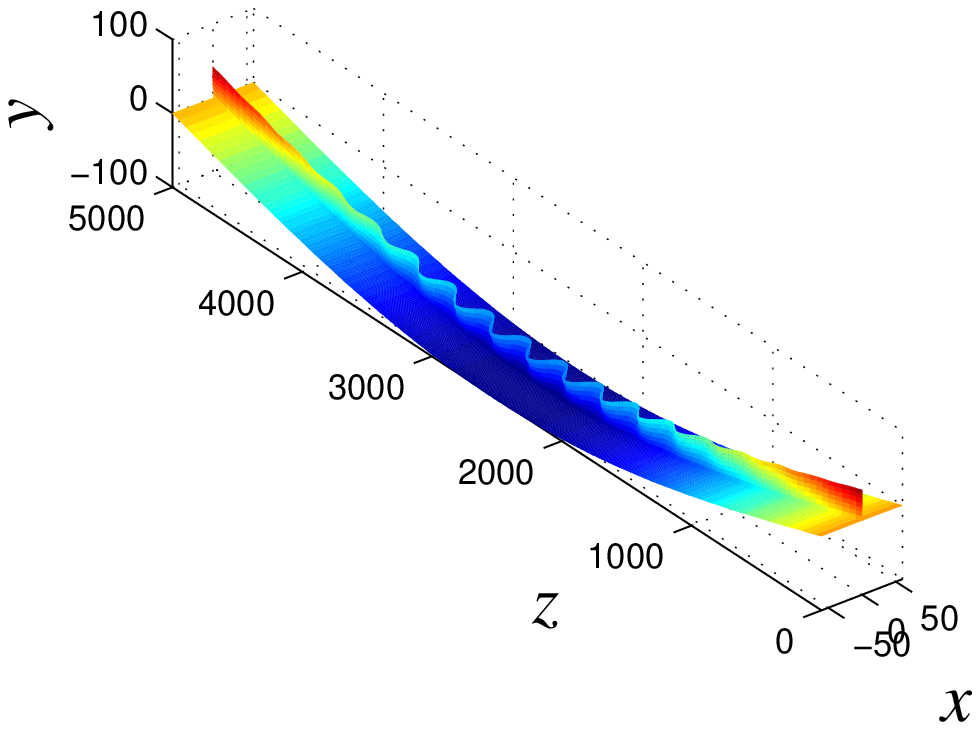}}
\caption{Numerical solutions from (left) \textsc{Abaqus} and (right)
  \textsc{Auto} visualized on 3-dimensional representations of the
  strut at the same load level: (a) $p=0.89$, (b) $p=0.84$, and (c)
  $p=0.81$ respectively).  All dimensions are in millimetres, but the
  local buckling displacements in the stiffener are scaled by a factor
  of 5 to aid visualization.}
\label{fig:3dab}
\end{figure}

Figure \ref{fig:abaqus}(d) shows the local out-of-plane deflection
profiles at the respective locations (i)--(iii) shown in Figure
\ref{fig:abaqus}(a--c), the comparisons being for the same values of
$q_s$. These, again, show excellent correlation between the results
from \textsc{Auto-07p} and \textsc{Abaqus}, particularly near the peak
load. A visual comparison between the 3-dimensional representation of
the strut from the analytical and the FE models is also
presented in Figure \ref{fig:3dab}.
The correlation with \textsc{Abaqus}, for a relatively high value of
$c_p$, is seen to be excellent, but when $c_p$ is decreased the
analytical model exhibits the characteristic snap-backs earlier (see
Figures \ref{fig:pg} and \ref{fig:cp_w}). However, the \textsc{Abaqus}
solution for static post-buckling, which relies on assuming an
initially imperfect geometry, does not (and indeed possibly cannot)
obtain the progressive cellular buckling solution. This is not
entirely surprising because the wavelength of the \textsc{Abaqus}
solution seems to be fixed throughout the loading history. This
demonstrates that for situations where buckling wavelengths are likely
to change, the conventional method for analysing instability problems
with static FE methods, where initial imperfections affine to a
combination of eigenmodes are introduced and analysed using a
nonlinear solver (e.g.\ the Riks method), may not be able to capture
some experimentally observed physical phenomena. As an alternative,
the authors attempted to model the problem as a dynamical system
within \textsc{Abaqus} such that the snap-backs may be replicated more
readily. However, computational issues in terms of slow and unreliable
convergence stalled this study and it has been left for the
future. Nevertheless, a recent study of a closely related structure
\cite{WB2013} clearly showed that the analytical approach matches the
physical response significantly more accurately than static FE
methods, in terms of the load versus displacement behaviour, and the
observed change in the post-buckling wavelength.  The observation of
precisely the cellular buckling response has been reported for the
flanges of thin-walled I-section struts experimentally
\cite{Becque_thesis}, which is very similar to the current case where
$h_1=h_2$ (see Figure \ref{fig:loadcross}). Hence, the authors are of
the opinion that since cellular buckling is an observed physical
phenomenon for those related structural components, the current
results from the analytical model are also valid.

\section{Concluding remarks}

The current work identifies a form of cellular buckling in axially
compressed stiffened plates. This arises from a potentially dangerous
interaction of local and global modes of buckling, with a
characteristic sequence of snap-back instabilities in the mechanical
response. It is also found that the effect of the instabilities can be
reduced somewhat by increasing the rotational stiffness of the joint
between the main plate and the stiffeners. However, the mode
interaction persists and the local buckling profile still changes
wavelength, as has been found in recent related studies
\cite{Becque2009a,WG2012,WB2013}. The increased rotational stiffness
of the joint allows the model to be validated from a purely numerical
formulation in FE software. However, earlier studies suggest that for
smaller values of the aforementioned stiffness, the analytical model
would be superior in modelling the actual physical response. This is
in terms of the equilibrium response, the load-carrying capacity and
the deformed shape, which is known from earlier research to exhibit
the change in the local buckling wavelength. The latter point is
crucial since the static FE solution seems to be unable to capture the
changing wavelength of the post-buckling mode that has been observed
in physical experiments. It is likely that in an actual scenario, the
subcritical response in conjunction with the snap-backs are likely to
lead to buckling cells being triggered dynamically. This has been
observed during the \emph{tripping} phenomenon \cite{Ronalds89} and
the postulate is that the elastic interactive buckling found
currently, if combined with plasticity, would promote this kind of
sudden and violent collapse of a stiffened panel.

Currently, the authors are extending the analytical model by including
the possibility of buckling the main plate in combination with the
stiffener. In this case, the load-carrying capacity is likely to be
reduced and it would be representative of a scenario where the
cross-section has uniform thickness. Moreover, an imperfection
sensitivity study is being conducted to quantify the parametric space
for designers to avoid such dangerous structural behaviour.

\bibliography{refs}

\end{document}